\documentclass[preprint,nofootinbib]{revtex4}
  \usepackage{amssymb,amsmath}
  \usepackage{graphicx}
  \begin{document}
  \title{Study of the $\overline{B}_{q}^{\ast}$ ${\to}$ $DM$ decays with perturbative QCD approach}
  \author{Junfeng Sun}
  \affiliation{Institute of Particle and Nuclear Physics,
              Henan Normal University, Xinxiang 453007, China}
  \author{Jie Gao}
  \affiliation{Institute of Particle and Nuclear Physics,
              Henan Normal University, Xinxiang 453007, China}
  \author{Yueling Yang}
  \affiliation{Institute of Particle and Nuclear Physics,
              Henan Normal University, Xinxiang 453007, China}
  \author{Qin Chang}
  \affiliation{Institute of Particle and Nuclear Physics,
              Henan Normal University, Xinxiang 453007, China}
  \author{Na Wang}
  \affiliation{Institute of Particle and Nuclear Physics,
              Henan Normal University, Xinxiang 453007, China}
  \author{Gongru Lu}
  \affiliation{Institute of Particle and Nuclear Physics,
              Henan Normal University, Xinxiang 453007, China}
  \author{Jinshu Huang}
  \affiliation{College of Physics and Electronic Engineering,
              Nanyang Normal University, Nanyang 473061, China}

  \begin{abstract}
  The $\overline{B}_{q}^{\ast}$ ${\to}$ $DP$, $DV$ weak decays
  are studied with the perturbative QCD approach, where $q$ $=$
  $u$, $d$ and $s$; $P$ and $V$ denote the ground $SU(3)$
  pseudoscalar and vector meson nonet.
  It is found that the branching ratios for the color-allowed
  $\overline{B}_{q}^{\ast}$ ${\to}$ $D_{q}{\rho}^{-}$ decays
  can reach up to $10^{-9}$ or more, and should be
  promisingly measurable at the running LHC and forthcoming
  SuperKEKB experiments in the near future.
  \end{abstract}
  \maketitle

  \section{Introduction}
  \label{sec01}
  In accordance with the conventional quark model assignments, the ground
  spin-singlet pseudoscalar $B_{q}$ mesons and spin-triplet vector
  $B^{\ast}_{q}$ mesons have the same flavor components, and consist
  of one valence heavy antiquark $\bar{b}$ and one light quark $q$,
  i.e., $\bar{b}q$, with $q$ $=$ $u$, $d$, $s$ \cite{pdg}.
  With the two $e^{+}e^{-}$ $B$-factory BaBar and Belle experiments,
  there is a combined data sample of over $1\,ab^{-1}$ at the
  ${\Upsilon}(4S)$ resonance. The $B_{u,d}$ meson weak decay modes with branching
  ratio of over $10^{-6}$ have been well measured \cite{epjc74}.
  The $B_{s}$ meson, which can be produced in hadron collisions or at/over
  the resonance ${\Upsilon}(5S)$ in $e^{+}e^{-}$ collisions\footnotemark[1],
  \footnotetext[1]{
  In hadron colliders, CDF and D0 each have accumulated about $10\,fb^{-1}$
  data, and LHCb has accumulated over $5\,fb^{-1}$ data up to the year
  of 2016 \cite{lhcb-operation}. In $e^{+}e^{-}$ colliders, Belle has
  accumulated more than $100\,fb^{-1}$ data at the resonance
  ${\Upsilon}(5S)$ \cite{epjc74}.}
  is being carefully scrutinized.
  However, the study of the $B_{q}^{\ast}$ mesons has not actually attracted
  much attention yet, subject to the relatively inadequate statistics.
  Because the mass of the $B_{q}^{\ast}$ mesons is a bit larger than
  that of the $B_{q}$ mesons, the $B_{q}^{\ast}$ meson should
  be produced at the relatively higher energy rather than at the
  resonance ${\Upsilon}(4S)$ in $e^{+}e^{-}$ collisions.
  With the high luminosities and large production cross section
  at the running LHC, the forthcoming SuperKEKB and future {\em Super
  proton proton Collider} (SppC, which is still in the preliminary
  discussion and research stage up to now), more and more $B_{q}^{\ast}$
  mesons will be accumulated in the future,
  which makes the $B_{q}^{\ast}$ mesons another research laboratory
  for testing the Cabibbo-Kobayashi-Maskawa (CKM) picture for
  $CP$-violating phenomena, examining our comprehension of the
  underlying dynamical mechanism for the weak decays of the heavy
  flavor hadrons.

  Having the same valence quark components and approximately an equal
  mass, both the $B^{\ast}_{q}$ and $B_{q}$ mesons can decay via
  weak interactions into the same final states.
  On the one hand, the $B^{\ast}_{q}$ and $B_{q}$ meson weak decays
  would provide each other with a spurious background;
  on the other hand, the interplay between the $B_{q}^{\ast}$ and $B_{q}$
  weak decays could offer some potential useful information to constrain
  parameters within the standard model, and might shed some fresh light on
  various intriguing puzzles in the $B_{q}$ meson decays.
  The $B_{q}$ meson decays are well described by the bottom quark decay
  with the light spectator quark $q$ in the spectator model.
  At the quark level, most of the hadronic $B_{q}$ meson decays involve
  the $b$ ${\to}$ $c$ transition due to the hierarchy relation among
  the CKM matrix elements.
  As is well known, there is a more than $3\,{\sigma}$ discrepancy between
  the value of ${\vert}V_{cb}{\vert}$ obtained from inclusive determinations,
  ${\vert}V_{cb}{\vert}$ $=$ $(42.2{\pm}0.8){\times}10^{-3}$, and from exclusive
  ones, ${\vert}V_{cb}{\vert}$ $=$ $(39.2{\pm}0.7){\times}10^{-3}$ \cite{pdg}.
  Besides the semileptonic $\overline{B}_{q}^{(\ast)}$ ${\to}$ $D^{(\ast)}{\ell}\bar{\nu}$
  decays, the nonleptonic $\overline{B}_{q}^{(\ast)}$ ${\to}$ $DM$
  decays, with $M$ representing the ground $SU(3)$ pseudoscalar $P$ and the
  vector $V$ meson nonet, are also induced by the $b$ ${\to}$ $c$ transition,
  and hence could be used to extract/constrain the CKM matrix element
  ${\vert}V_{cb}{\vert}$.

  From the dynamical point of view, the phenomenological models used for the
  $\overline{B}_{q}$ ${\to}$ $DM$ decays might, in principle, be extended
  and applied to the $\overline{B}_{q}^{\ast}$ ${\to}$ $DM$ decays.
  The practical applicability and reliability of these models could be
  reevaluated with the $\overline{B}_{q}^{\ast}$ ${\to}$ $DM$ decays.
  Recently, some attractive QCD-inspired methods, such as the perturbative
  QCD (pQCD) approach \cite{prd52.3958,prd55.5577,prd56.1615,plb504.6,
  prd63.054008,prd63.074006,prd63.074009,epjc23.275},
  the QCD factorization (QCDF) approach \cite{prl83.1914,npb591.313,
  npb606.245,plb488.46,plb509.263,prd64.014036,npb774.64,prd77.074013},
  soft and collinear effective theory \cite{prd63.014006,prd63.114020,
  plb516.134,prd65.054022,prd66.014017,npb643.431,plb553.267,npb685.249}
  and so on, have been developed vigorously and employed widely to
  explain measurements on the $B_{q}$ meson decays.
  The $\overline{B}_{q}$ ${\to}$ $DM$ decays have been studied with
  the QCDF \cite{npb591.313,plb476.339} and pQCD \cite{prd69.094018,prd78.014018}
  approaches, but there are few research works on the $B_{q}^{\ast}$ meson
  weak decays.
  Recently, the $\overline{B}_{q}^{\ast}$ ${\to}$ $D_{q}V$ decays have been
  investigated with the QCDF approach \cite{epjc76.523}, and it is shown
  that the $\overline{B}_{q}^{{\ast}0}$ ${\to}$ $D_{q}^{+}{\rho}^{-}$ decays
  with branching ratios of ${\cal O}(10^{-8})$ might be accessible to the
  existing and future heavy flavor experiments.
  In this paper, we will give a comprehensive investigation into the
  two-body nonleptonic $\overline{B}_{q}^{\ast}$ ${\to}$ $DM$ decays
  with the pQCD approach in order to provide the future experimental
  research with an available reference.

  As is well known, the $B^{\ast}_{q}$ meson decays are dominated by
  the electromagnetic interactions rather than the weak interactions,
  which differs significantly from the $B_{q}$ meson decays.
  One can easily expect that the branching ratios for the
  $\overline{B}_{q}^{\ast}$ ${\to}$ $DM$ weak decays should be very
  small due to the short electromagnetic lifetimes of the $B_{q}^{\ast}$
  mesons \cite{epja52.90}, although these processes are
  favored by the CKM matrix element ${\vert}V_{cb}{\vert}$.
  Of course, an abnormal large branching ratio might be a possible hint
  of new physics beyond the standard model.
  There is still no experimental report on the $\overline{B}_{q}^{\ast}$
  ${\to}$ $DM$ weak decays so far.
  Furthermore, the $\overline{B}_{q}^{\ast}$ ${\to}$ $DM$ weak decays
  offer the unique opportunity of observing the weak decay of a vector
  meson, where polarization effects could be explored.

  This paper is organized as follows.
  In section \ref{sec02}, we present the theoretical framework, the
  conventions and notations, together with amplitudes for the
  $\overline{B}_{q}^{\ast}$ ${\to}$ $DM$ decays.
  Section \ref{sec03} is devoted to the numerical results and discussion.
  The final section is a summary.

  \section{theoretical framework}
  \label{sec02}
  \subsection{The effective Hamiltonian}
  \label{sec0201}
  As is well known, the weak decays of the $B_{q}^{(\ast)}$ mesons
  inevitably involve multiple length scales, including the mass of
  $m_{W}$ for the virtual gauge boson $W$, the mass of $m_{b}$ for
  the decaying bottom quark, the infrared confinement scale
  ${\Lambda}_{\rm QCD}$ of the strong interactions, and
  $m_{W}$ ${\gg}$ $m_{b}$ ${\gg}$ ${\Lambda}_{\rm QCD}$.
  So, one usually has to resort to the effective theory approximation
  scheme. With the operator product expansion and the renormalization
  group (RG) method, the effective Hamiltonian for the
  $\overline{B}_{q}^{\ast}$ ${\to}$ $DM$ decays can be written as
  \cite{9512380},
   \begin{equation}
  {\cal H}_{\rm eff}\, =\, \frac{G_{F}}{\sqrt{2}}\,
   \sum\limits_{q^{\prime}=d,s} V_{cb}\,V_{uq^{\prime}}^{\ast}
   \Big\{ C_{1}({\mu})\,Q_{1}({\mu})+C_{2}({\mu})\,Q_{2}({\mu})\Big\}
  +{\rm h.c.}
   \label{hamilton},
   \end{equation}
  where $G_{F}$ ${\simeq}$ $1.166{\times}10^{-5}\,{\rm GeV}^{-2}$
  \cite{pdg} is the Fermi coupling constant.

  Using the Wolfenstein parametrization, the CKM factor
  $V_{cb}V_{uq^{\prime}}^{\ast}$ are expressed as a series expansion of
  the small Wolfenstein parameter ${\lambda}$ ${\approx}$ $0.2$ \cite{pdg}.
  Up to the order of ${\cal O}({\lambda}^{7})$, they
  can be written as follows:
  \begin{eqnarray}
  V_{cb}\,V_{ud}^{\ast}
  &=& A\,{\lambda}^{2}\,( 1 -{\lambda}^{2}/2-{\lambda}^{4}/8 )
      +{\cal O}({\lambda}^{7})
  \label{vcbvud}, \\
  V_{cb}\,V_{us}^{\ast}
  &=& A\,{\lambda}^{3}+{\cal O}({\lambda}^{7})
  \label{vcbvus}.
  \end{eqnarray}
  It is very clearly seen that the both $V_{cb}\,V_{ud}^{\ast}$
  and $V_{cb}\,V_{us}^{\ast}$ are real-valued, i.e.,
  there is no weak phase difference.
  However, nonzero weak phase difference is necessary and
  indispensable for the direct $CP$ violation.
  Therefore, none of direct $CP$ violation should be expected
  for the $\overline{B}_{q}^{\ast}$ ${\to}$ $DM$ decays.

  The renormalization scale ${\mu}$ separates the physical
  contributions into the short- and long-distance parts.
  The Wilson coefficients $C_{1,2}$ summarize the physical
  contributions above the scale ${\mu}$.
  They, in principle, are calculable order by order in the
  strong coupling ${\alpha}_{s}$ at the scale $m_{W}$ with
  the ordinary perturbation theory,
  and then evolved with the RG equation to the characteristic
  scale ${\mu}$ ${\sim}$ ${\cal O}(m_{b})$ for the bottom quark
  decay \cite{9512380}.
  The Wilson coefficients at the scale $m_{W}$ are determined at the
  quark level rather than the hadron level, so they are regarded
  as process-independent couplings of the local operators $Q_{i}$.
  Their explicit analytical expressions, including the next-to-leading
  order corrections, have been given in Ref.\cite{9512380}.

  The physical contributions from the scales lower than ${\mu}$ are
  contained in the hadronic matrix elements (HME) where the local
  four-quark operators are sandwiched between the initial and final
  hadron states.
  The local six-dimension operators arising from the $W$-boson exchange
  are defined as follows:
  \begin{eqnarray}
  Q_{1} &=& [\,\bar{c}_{\alpha}\,{\gamma}_{\mu}\,(1-{\gamma}_{5})\,b_{\alpha}]\,\
            [\,\bar{q}^{\prime}_{\beta}\,{\gamma}^{\mu}\,(1-{\gamma}_{5})\,u_{\beta}]
  \label{operator:q1},
  \\
  Q_{2} &=& [\,\bar{c}_{\alpha}\,{\gamma}_{\mu}\,(1-{\gamma}_{5})\,b_{\beta}]\,\
            [\,\bar{q}^{\prime}_{\beta}\,{\gamma}^{\mu}\,(1-{\gamma}_{5})\,u_{\alpha}]
  \label{operator:q2}.
  \end{eqnarray}
  where ${\alpha}$ and ${\beta}$ are color indices, i.e., the gluonic
  corrections are included.
  The operator $Q_{1}$ ($Q_{2}$) consists of two color-singlet
  (color-octet) currents.
  The operators $Q_{1}$ and $Q_{2}$, called current-current operators
  or tree operators, have the same flavor form and a different color structure.
  It is obvious that the $\overline{B}_{q}^{\ast}$ ${\to}$ $DM$ decays
  are uncontaminated by the contributions from the penguin operators,
  which is positive to extract the CKM matrix element ${\vert}V_{cb}{\vert}$.

  Because of the participation of the strong interaction, especially, the
  long-distance effects in the conversion from the quarks of the local
  operators to the initial and final hadrons, barricades are still erected
  on the approaches of nonleptonic $\overline{B}_{q}^{(\ast)}$ weak
  decays, which complicates the calculation.
  HME of the local operators are the most intricate part for theoretical
  calculation, where the perturbative and nonperturbative contributions
  entangle with each other.
  To evaluate the HME amplitudes, one usually has to resort to some
  plausible approximations and assumptions, which results in the
  model-dependence of theoretical predictions.
  It is obvious that a large part of the uncertainties does come from
  the practical treatment of HME, due to our inadequate understanding
  of the hadronization mechanism and the low-energy QCD behavior.
  For the phenomenology of the $\overline{B}_{q}^{\ast}$ ${\to}$ $DM$
  decays, one of the main tasks at this stage is how to effectively
  factorize HME of the local operators into hard and soft parts, and
  how to evaluate HME properly.

  \subsection{Hadronic matrix elements}
  \label{sec0202}
  One of the phenomenological schemes for the HME calculation is the
  factorization approximation based on Bjorken's {\em a priori} color
  transparency hypothesis, which says that the color singlet energetic
  hadron would have flown rapidly away from the color fields existing
  in the neighborhood of the interaction point before the soft gluons
  are exchanged among hadrons \cite{npb11.325}.
  Modeled on the amplitudes for exclusive processes with the Lepage-Brodsky
  approach \cite{prd22.2157}, HME are usually written as the convolution integral
  of the hard kernels and the hadron distribution amplitudes (DAs).
  Hard kernels are expressed as the scattering amplitudes for the transition
  of the heavy bottom quark into light quarks.
  They are generally computable at the quark level with the perturbation
  theory as a series of expansion in the parameter $1/m_{b}$ and the strong
  coupling constant ${\alpha}_{s}$ in the heavy quark limit.
  It is assumed that the soft and nonperturbative contributions of HME
  could be absorbed into hadron DAs.
  The distribution amplitudes are functions of parton momentum fractions.
  They, although not calculable, are regarded as universal
  and can be determined by nonperturbative means or extracted from data.
  With the traits of universality and determinability of hadron DAs, HME
  have a sample structure and can be evaluated to make predictions.

  Besides the factorizable contributions to HME, the nonfactorizable
  corrections to HME also play an important role in commenting on
  the experimental measurements and solving the so-called puzzles and
  anomalies, and hence should be carefully considered, as commonly
  recognized by theoretical physicists.
  In order to regulate the endpoint singularities which appear
  in the spectator scattering and annihilation amplitudes with the
  QCDF approach and spoil the perturbative calculation with the
  collinear approximation \cite{npb591.313,npb606.245,plb488.46,
  plb509.263,prd64.014036}, it is suggested by the pQCD approach
  \cite{prd52.3958,prd55.5577,prd56.1615,plb504.6,prd63.054008,
  prd63.074006,prd63.074009,epjc23.275} that the transverse momentum
  of quarks should be conserved and, additionally, that a Sudakov factor
  should be introduced to DAs for all the participant
  hadrons to further suppress the long-distance and soft
  contributions.
  The basic pQCD formula for nonleptonic weak decay amplitudes
  could be factorized into three parts: the hard effects enclosed
  by the Wilson coefficients $C_{i}$, hard scattering kernels
  ${\cal H}_{i}$, and the universal wave functions ${\Phi}_{j}$.
  The general form is a multidimensional integral
  \cite{prd52.3958,prd55.5577,prd56.1615,plb504.6,prd63.054008,
  prd63.074006,prd63.074009,epjc23.275},
  \begin{equation}
 {\cal A}_{i}\ {\propto}\ {\int}\, {\prod_j}dx_{j}\,db_{j}\,
  C_{i}(t)\,{\cal H}_{i}(t_{i},x_{j},b_{j})\,{\Phi}_{j}(x_{j},b_{j})\,e^{-S_{j}}
  \label{hadronic},
  \end{equation}
  where $x_{j}$ is the longitudinal momentum fraction of the valence
  quarks. $b_{j}$ is the conjugate variable of the transverse momentum $k_{jT}$.
  The scale $t_{i}$ is preferably chosen to be the maximum virtuality of
  all the internal particles. The Sudakov factor $e^{-S_{j}}$, together
  with the particular scale $t_{i}$, will ensure the perturbative
  calculation is feasible and reliable.

  \subsection{Kinematic variables}
  \label{sec0203}

  The $\overline{B}_{q}^{(\ast)}$ weak decays are actually dominated
  by the $b$ quark weak decay.
  In the heavy quark limit, the light quark originating from the heavy
  bottom quark decay is assumed to be energetic and race quickly away
  from the weak interaction point.
  If the velocity $v$ ${\sim}$ $c$ (the speed of light), the light
  quarks move near the light-cone line.
  The light-cone dynamics can be used to describe the relativistic
  system along the light-front direction.
  The light-cone coordinates $(x^{+},x^{-},x_{\perp})$ of space-time
  are defined as $x^{\pm}$
  $=$ $(x^{0}{\pm}x^{3})/\sqrt{2}$ (or $(t{\pm}x^{3})/\sqrt{2}$) and
  $x_{\perp}$ $=$ $x^{i}$ with $i$ $=$ $1$ and $2$.
  $x^{\pm}$ $=$ $0$ is called the light-front.
  The scalar product of any two four-dimensional vectors is given by
  $a{\cdot}b$ $=$ $a_{\mu}b^{\mu}$ $=$ $a^{+}b^{-}$ $+$ $a^{-}b^{+}$
  $-$ $a_{\perp}{\cdot}b_{\perp}$.
  In the rest frame of the $\overline{B}_{q}^{\ast}$ meson, the final
  $D$ and $M$ mesons are back-to-back.
  The light-cone kinematic variables are defined as follows:
  \begin{equation}
  p_{\overline{B}_{q}^{\ast}}\, =\, p_{1}\, =\, \frac{m_{1}}{\sqrt{2}}(1,1,0)
  \label{kine-p1},
  \end{equation}
  \begin{equation}
  p_{D}\, =\, p_{2}\, =\, (p_{2}^{+},p_{2}^{-},0)
  \label{kine-p2},
  \end{equation}
  \begin{equation}
  p_{M}\, =\, p_{3}\, =\, (p_{3}^{-},p_{3}^{+},0)
  \label{kine-p3},
  \end{equation}
  \begin{equation}
  k_{i}\, =\, x_{i}\,p_{i}+(0,0,k_{iT})
  \label{kine-ki},
  \end{equation}
  \begin{equation}
  p_{i}^{\pm}\, =\, (E_{i}\,{\pm}\,p)/\sqrt{2}
  \label{kine-pipm},
  \end{equation}
  \begin{equation}
  t\, =\, 2\,p_{1}{\cdot}p_{2}\, =\, m_{1}^{2}+m_{2}^{2}-m_{3}^{2}\, =\,2\,m_{1}\,E_{2}
  \label{kine-t},
  \end{equation}
  \begin{equation}
  u\, =\, 2\,p_{1}{\cdot}p_{3}\, =\, m_{1}^{2}-m_{2}^{2}+m_{3}^{2}\, =\,2\,m_{1}\,E_{3}
  \label{kine-u},
  \end{equation}
  \begin{equation}
  s\, =\, 2\,p_{2}{\cdot}p_{3}\, =\, m_{1}^{2}-m_{2}^{2}-m_{3}^{2}
  \label{kine-s},
  \end{equation}
  \begin{equation}
  s\,t +s\,u-t\,u \,=\, 4\,m_{1}^{2}\,p^{2}
  \label{kine-pcm},
  \end{equation}
  where the subscripts $i$ $=$ $1$, $2$ and $3$ of the variables (such as,
  four-dimensional momentum $p_{i}$, energy $E_{i}$, and mass $m_{i}$)
  correspond to the $\overline{B}_{q}^{\ast}$, $D$ and $M$ mesons,
  respectively.
  $k_{i}$ is the momentum of the light antiquark carrying the longitudinal
  momentum fraction $x_{i}$. $k_{iT}$ is the transverse momentum.
  $t$, $u$ and $s$ are the Lorentz scalar variables. $p$ is the
  common momentum of the final states.
  These momenta are shown in Fig.\ref{fig:fey-t}(a), Fig.\ref{fig:fey-c}(a)
  and Fig.\ref{fig:fey-a}(a).

  \subsection{Wave functions}
  \label{sec0204}
  As aforementioned, wave functions are the essential input parameters
  in the master pQCD formula for the HME calculation.
  Following the notations in Refs. \cite{npb529.323,prd65.014007,prd92.074028,
  plb751.171,plb752.322,jhep9901.010,jhep0703.069,jhep0605.004}, the wave functions
  of the participating meson are defined as the meson-to-vacuum HME.
  \begin{equation}
 {\langle}0{\vert}\bar{q}_{i}(z)b_{j}(0){\vert}
  \overline{B}_{q}^{\ast}(p,{\epsilon}^{\parallel}){\rangle}\,
 =\, \frac{f_{B_{q}^{\ast}}}{4} {\int}d^{4}k\,e^{-ik{\cdot}z}
  \Big\{ \!\!\not{\!\epsilon}^{\parallel}\, \Big[
  m_{B_{q}^{\ast}}\, {\Phi}_{B_{q}^{\ast}}^{v}(k)\, -
  \!\not{p}\, {\Phi}_{B_{q}^{\ast}}^{t}(k) \Big] \Big\}_{ji}
  \label{wf-bq01},
  \end{equation}
  \begin{equation}
 {\langle}0{\vert}\bar{q}_{i}(z)b_{j}(0){\vert}
  \overline{B}_{q}^{\ast}(p,{\epsilon}^{\perp}){\rangle}\,
 =\, \frac{f_{B_{q}^{\ast}}}{4} {\int}d^{4}k\,e^{-ik{\cdot}z}
  \Big\{ \!\!\not{\!\epsilon}^{\perp}\, \Big[
  m_{B_{q}^{\ast}}\, {\Phi}_{B_{q}^{\ast}}^{V}(k)\, -
  \!\not{p}\, {\Phi}_{B_{q}^{\ast}}^{T}(k) \Big] \Big\}_{ji}
  \label{wf-bc02},
  \end{equation}
  \begin{equation}
 {\langle}D_{q}(p){\vert}\bar{c}_{i}(0)q_{j}(z){\vert}0{\rangle}\,
 =\, \frac{i\,f_{D_{q}}}{4}{\int}d^{4}k\,e^{+ik{\cdot}z}\,
  \Big\{ {\gamma}_{5}\Big[ \!\!\not{p}\,{\Phi}_{D_{q}}^{a}(k)
  +m_{D_{q}}\,{\Phi}_{D_{q}}^{p}(k) \Big] \Big\}_{ji}
  \label{wf-cq01},
  \end{equation}
  \begin{eqnarray}
 {\langle}P(p){\vert}\bar{q}_{i}(0)q^{\prime}_{j}(z){\vert}0{\rangle}\,
 &=& \frac{1}{4}{\int}d^{4}k\,e^{+ik{\cdot}z}\,
  \Big\{ {\gamma}_{5}\Big[ \!\!\not{p}\,{\Phi}_{P}^{a}(k) +
 {\mu}_{P}\,{\Phi}_{P}^{p}(k)
  \nonumber \\ & & \qquad \qquad \qquad +
 {\mu}_{P}\,(\not{n}_{+}\!\!\not{n}_{-}-1)\,{\Phi}_{P}^{t}(k)
  \Big] \Big\}_{ji}
  \label{wf-p},
  \end{eqnarray}
  \begin{equation}
 {\langle}V(p,{\epsilon}^{\parallel}){\vert}\bar{q}_{i}(0)q^{\prime}_{j}(z){\vert}0{\rangle}\,
 =\, \frac{1}{4}{\int}d^{4}k\,e^{+ik{\cdot}z}\,
  \Big\{ \!\!\not{\epsilon}^{\parallel}\,m_{V}\,{\Phi}_{V}^{v}(k)
  +\!\!\not{\epsilon}^{\parallel}\!\!\not{p}\,{\Phi}_{V}^{t}(k)
  -m_{V}\,{\Phi}_{V}^{s}(k) \Big\}_{ji}
  \label{wf-v-el},
  \end{equation}
  \begin{eqnarray}
 {\langle}V(p,{\epsilon}^{\perp}){\vert}\bar{q}_{i}(0)q^{\prime}_{j}(z){\vert}0{\rangle}\,
 &=& \frac{1}{4}{\int}d^{4}k\,e^{+ik{\cdot}z}\,
  \Big\{ \!\!\not{\epsilon}^{\perp}\,m_{V}\,{\Phi}_{V}^{V}(k)
  +\!\!\not{\epsilon}^{\perp}\!\!\not{p}\,{\Phi}_{V}^{T}(k)
  \nonumber \\ & & \qquad +
  \frac{i\,m_{V}}{p{\cdot}n_{+}}\,{\gamma}_{5}\,{\varepsilon}_{{\mu}{\nu}{\alpha}{\beta}}
  {\gamma}^{\mu}\,{\epsilon}^{{\perp}{\nu}}\,p^{\alpha}\,
  n_{+}^{\beta}\,{\Phi}_{V}^{A}(k) \Big\}_{ji}
  \label{wfv-et},
  \end{eqnarray}
  where $f_{B_{q}^{\ast}}$ and $f_{D_{q}}$ are the decay
  constants of the $\overline{B}_{q}^{\ast}$ meson and the $D_{q}$
  meson, respectively.
  ${\epsilon}^{\parallel}$ and ${\epsilon}^{\perp}$ are the
  longitudinal and transverse polarization vectors.
  $n_{+}$ $=$ $(1,0,0)$ and $n_{-}$ $=$ $(0,1,0)$
  are the positive and negative null vectors,
  i.e., $n_{\pm}^{2}$ $=$ $0$. The chiral factor ${\mu}_{P}$
  relates the pseudoscalar meson mass to the quark mass through
  the following way \cite{jhep9901.010},
  \begin{equation}
 {\mu}_{P}\, =\, \frac{m_{\pi}^{2}}{m_{u}+m_{d}}
          \, =\, \frac{m_{K}^{2}}{m_{u,d}+m_{s}}
          \, {\approx}\, (1.6{\pm}0.2)\,\text{GeV}
  \label{up}.
  \end{equation}

  With the twist classification based on the power counting rule
  in the infinite momentum frame \cite{npb529.323,prd65.014007},
  the wave functions ${\Phi}_{B_{q}^{\ast},V}^{v,T}$ and
  ${\Phi}_{D_{q},P}^{a}$ are twist-2 (the leading twist),
  while the wave functions ${\Phi}_{B_{q}^{\ast},V}^{t,V,s,A}$
  and ${\Phi}_{D_{q},P}^{p,t}$ are twist-3.
  By integrating out the transverse momentum from the wave
  functions, one can obtain the corresponding distribution
  amplitudes.
  In our calculation, the expressions of the DAs for the
  heavy-flavored mesons are \cite{prd92.074028,plb751.171,plb752.322}
   \begin{equation}
  {\phi}_{B_{q}^{\ast}}^{v,T}(x) = A\, x\,\bar{x}\,
  {\exp}\Big\{ -\frac{1}{8\,{\omega}_{B_{q}^{\ast}}^{2}}\,
   \Big( \frac{m_{q}^{2}}{x}+\frac{m_{b}^{2}}{\bar{x}} \Big) \Big\}
   \label{da-bqlv},
   \end{equation}
   \begin{equation}
  {\phi}_{B_{q}^{\ast}}^{t}(x) = B\, (\bar{x}-x)^{2}\,
  {\exp}\Big\{ -\frac{1}{8\,{\omega}_{B_{q}^{\ast}}^{2}}\,
   \Big( \frac{m_{q}^{2}}{x}+\frac{m_{b}^{2}}{\bar{x}} \Big) \Big\}
   \label{da-bqlt},
   \end{equation}
   \begin{equation}
  {\phi}_{B_{q}^{\ast}}^{V}(x) = C\, \{1+(\bar{x}-x)^{2}\}\,
  {\exp}\Big\{ -\frac{1}{8\,{\omega}_{B_{q}^{\ast}}^{2}}\,
   \Big( \frac{m_{q}^{2}}{x}+\frac{m_{b}^{2}}{\bar{x}} \Big) \Big\}
   \label{da-bqtt},
   \end{equation}
   \begin{equation}
  {\phi}_{D_{q}}^{a}(x) = D\, x\,\bar{x}\,
  {\exp}\Big\{ -\frac{1}{8\,{\omega}_{D_{q}}^{2}}\,
   \Big( \frac{m_{q}^{2}}{x}+\frac{m_{c}^{2}}{\bar{x}} \Big) \Big\}
   \label{da-cqa},
   \end{equation}
   \begin{equation}
  {\phi}_{D_{q}}^{p}(x) = E\,
  {\exp}\Big\{ -\frac{1}{8\,{\omega}_{D_{q}}^{2}}\,
   \Big( \frac{m_{q}^{2}}{x}+\frac{m_{c}^{2}}{\bar{x}} \Big) \Big\}
   \label{da-cqp},
   \end{equation}
   where $x$ and $\bar{x}$ (${\equiv}$ $1$ $-$ $x$) are the
   longitudinal momentum fractions of the light and heavy partons;
   $m_{b}$, $m_{c}$ and $m_{q}$ are the mass of the valence
   $b$, $c$ and $q$ quarks.
   The parameter ${\omega}_{i}$ determines the average transverse
   momentum of the partons, and
   ${\omega}_{i}$ ${\approx}$ $m_{i}\,{\alpha}_{s}(m_{i})$.
   The parameters $A$, $B$, $C$, $D$ and $E$ are the normalization
   coefficients to satisfy the conditions,
   \begin{equation}
  {\int}_{0}^{1}dx\,{\phi}_{B_{q}^{\ast}}^{v,t,V,T}(x)=1
   \label{wave-nb},
   \end{equation}
   \begin{equation}
  {\int}_{0}^{1}dx\,{\phi}_{D_{q}}^{a,p}(x) =1
   \label{wave-nd}.
   \end{equation}

  The main distinguishing feature of the above DAs in
  Eqs.(\ref{da-bqlv}-\ref{da-cqp}) is the exponential
  functions, where the exponential factors are proportional
  to the ratio of the parton mass squared $m_{i}^{2}$ to the
  momentum fraction $x_{i}$, i.e., $m_{i}^{2}/x_{i}$.
  Hence, the DAs of Eqs.(\ref{da-bqlv}-\ref{da-cqp}) are generally
  consistent with the ansatz that the momentum fractions are
  shared among the valence quarks according to the quark mass,
  i.e., a light quark will carry a smaller fraction of the parton
  momentum than a heavy quark in a heavy-light system.
  In addition, the exponential functions strongly suppress
  the contributions from the endpoint of $x$, $\bar{x}$ ${\to}$
  $0$, and naturally provide the effective truncation for the
  endpoint and soft contributions.

  As is well known, there are many phenomenological DA models for
  the charmed mesons. Some have been recited by Eq.(30) in
  Ref.\cite{prd78.014018}. One of the favorable DA models from
  the experimental data, without the distinction between the
  twist-2 and twist-3, has the common expression as below,
   \begin{equation}
  {\phi}_{D_{q}}(x) = 6\,x\,\bar{x}\,\big\{1+C_{D_{q}}\,(\bar{x}-x) \big\}
   \label{wave-d-xb},
   \end{equation}
  where the parameter $C_{D_{u,d}}$ $=$ $0.5$ for the $D_{u,d}$ meson,
  and $C_{D_{s}}$ $=$ $0.4$ for the $D_{s}$ meson.

  The expressions of the twist-2 quark-antiquark DAs for the
  light pseudoscalar and vector mesons have the expansion
  \cite{jhep9901.010,jhep0703.069,jhep0605.004},
   \begin{equation}
  {\phi}_{P}^{a}(x)\, =\, i\,f_{P}\,6\,x\,\bar{x}\,
   \sum\limits_{i=0} a^{P}_{i}\,C_{i}^{3/2}({\xi})
   \label{da-pa},
   \end{equation}
   \begin{equation}
 {\phi}_{V}^{v}(x) \, =\, f_{V}\,6\,x\,\bar{x}\,
  \sum\limits_{i=0} a^{\parallel}_{i}\,C_{i}^{3/2}({\xi})
  \label{da-rho-v},
  \end{equation}
  \begin{equation}
 {\phi}_{V}^{T}(x) \, =\, f_{V}^{T}\,6\,x\,\bar{x}\,
  \sum\limits_{i=0} a^{\perp}_{i}\,C_{i}^{3/2}({\xi})
  \label{da-rho-T},
  \end{equation}
  where $f_{P}$ is the decay constant for the pseudoscalar meson $P$;
  $f_{V}$ and $f_{V}^{T}$ are the vector and tensor (also called the
  longitudinal and transverse) decay constants for the vector meson $V$.
  The nonperturbative parameters of $a_{i}^{P,{\parallel},{\perp}}$
  are called the Gegenbauer moments, and $a_{0}^{P,{\parallel},{\perp}}$
  $=$ $1$ for the asymptotic forms, $a_{{\rm odd}~i}^{P,{\parallel},{\perp}}$
  $=$ $0$ for the DAs of the $G$-parity eigenstates, such as the unflavored
  ${\pi}$, ${\eta}$, ${\eta}^{\prime}$, ${\rho}$, ${\omega}$, ${\phi}$ mesons.
  The short-hand notation ${\xi}$ $=$ $x$ $-$ $\bar{x}$ $=$ $2\,x$ $-$ $1$.
  The analytical expressions of the Gegenbauer polynomials $C_{i}^{j}({\xi})$
  are as below,
   \begin{eqnarray}
   & & C_{0}^{j}({\xi})\, =\, 1
   \label{eq-c0}, \\
   & & C_{1}^{j}({\xi})\, =\, 2\,j\,{\xi}
   \label{eq-c1}, \\
   & & C_{2}^{j}({\xi})\, =\, 2\,j\,(j+1)\,{\xi}^{2}-j
   \label{eq-c2}, \\
   & & ...... \nonumber
   \end{eqnarray}

  As for the twist-3 DAs for the light pseudoscalar and vector mesons,
  their asymptotic forms will be employed in this paper for the
  simplification \cite{prd78.014018,jhep9901.010,jhep0703.069,jhep0605.004}, i.e.,
   \begin{eqnarray}
  {\phi}_{P}^{p}(x)&=& +i\,f_{P}\, C_{0}^{1/2}({\xi}) 
   \label{da-pp}, \\
  {\phi}_{P}^{t}(x)&=& -i\,f_{P}\, C_{1}^{1/2}({\xi})
   \label{da-pt}, \\
 {\phi}_{V}^{t}(x) &=& +3\, f_{V}^{T}\,{\xi}^{2}
  \label{da-rho-t}, \\
 {\phi}_{V}^{s}(x) &=& -3\, f_{V}^{T}\,{\xi}
  \label{da-rho-s}, \\
 {\phi}_{V}^{V}(x) &=& +\frac{3}{4}\,f_{V}\,(1+{\xi}^{2})
  \label{da-rho-V}, \\
 {\phi}_{V}^{A}(x) &=& -\frac{3}{2}\,f_{V}\,{\xi}
  \label{da-rho-A}.
  \end{eqnarray}

  \subsection{Decay amplitudes}
  \label{sec0205}
\begin{figure}[h]
  \includegraphics[width=0.90\textwidth,bb=90 625 515 720]{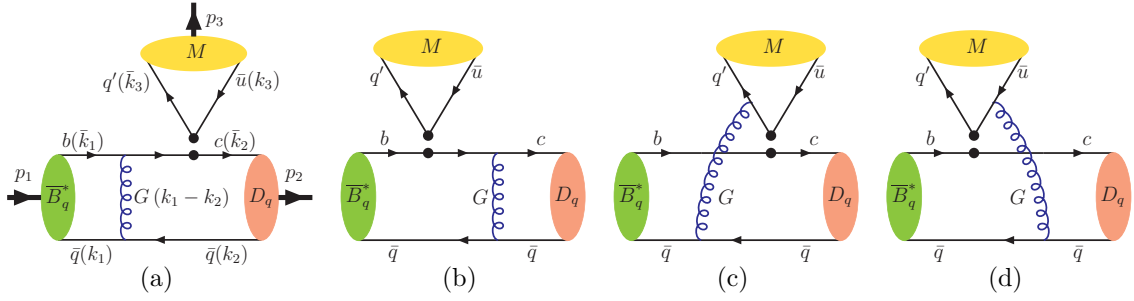}
  \caption{The color-allowed diagrams for the $\overline{B}_{q}^{\ast}$
   ${\to}$ $D_{q}M$ decays with the pQCD approach, where (a,b) and
   (c,d) are factorizable and nonfactorizable emission topologies,
   respectively.}
  \label{fig:fey-t}
  \end{figure}

\begin{figure}[h]
  \includegraphics[width=0.90\textwidth,bb=90 625 515 710]{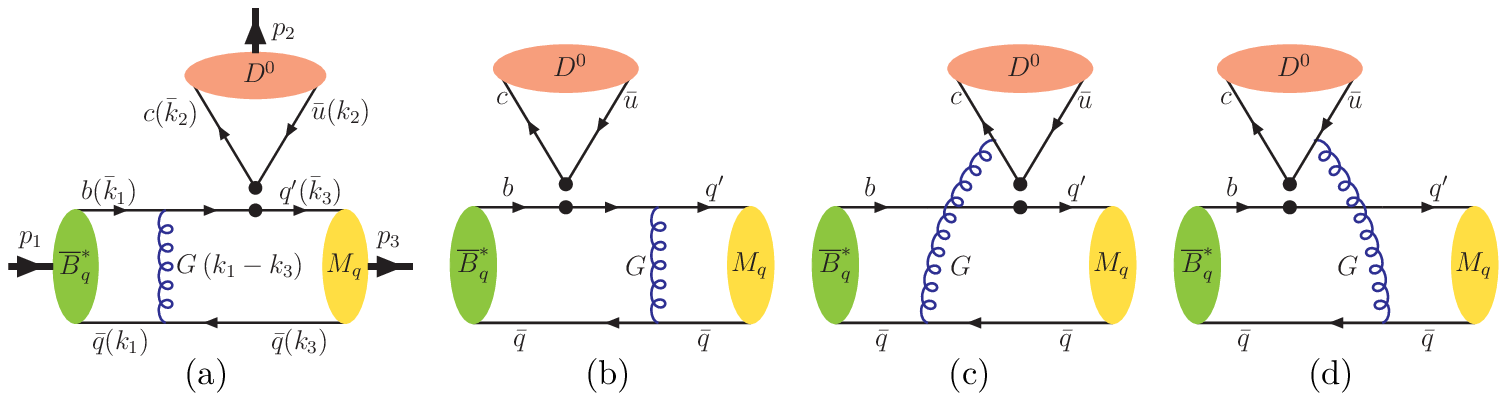}
  \caption{The color-suppressed diagrams for the $\overline{B}^{\ast}_{q}$
   ${\to}$ $D^{0}M_{q}$ decays with the pQCD approach.}
  \label{fig:fey-c}
  \end{figure}

\begin{figure}[h]
  \includegraphics[width=0.90\textwidth,bb=90 625 515 710]{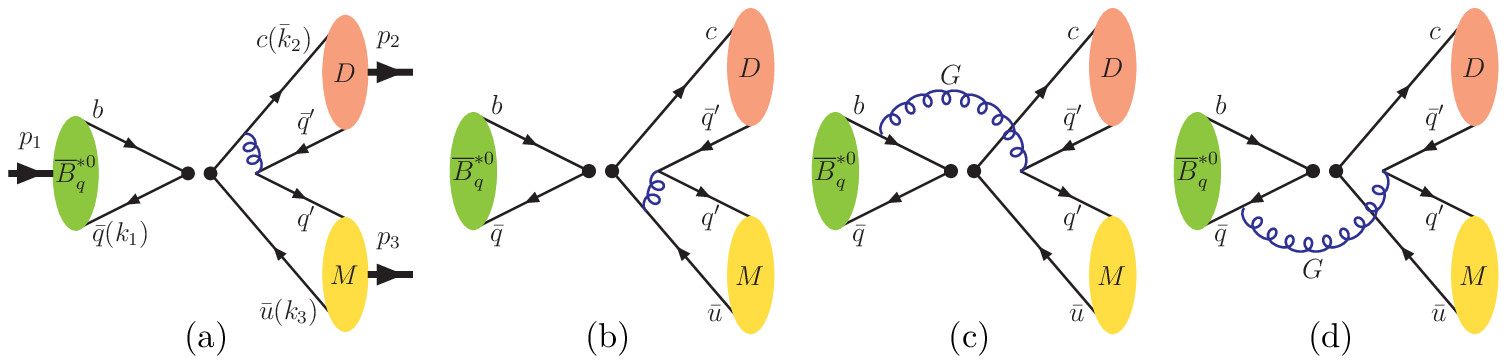}
  \caption{The annihilation diagrams for the $\overline{B}_{q}^{{\ast}0}$
   ${\to}$ $DM$ decays with the pQCD approach.}
  \label{fig:fey-a}
  \end{figure}

  As aforementioned, the $\overline{B}_{q}^{\ast}$ ${\to}$ $DM$ weak decays
  are induced practically by the $b$ quark decay at the quark level.
  There are three possible types of Feynman diagrams for the
  $\overline{B}_{q}^{\ast}$ ${\to}$ $DM$ decays with the pQCD approach, i.e.,
  the color-allowed topologies of Fig.\ref{fig:fey-t} induced by
  the external $W$-emission interactions,
  the color-suppressed topologies of Fig.\ref{fig:fey-c} induced by
  the internal $W$-emission interactions,
  and the annihilation topologies of Fig.\ref{fig:fey-a} induced by
  the $W$-exchange interactions.
  In the emission topologies of Fig.\ref{fig:fey-t} (Fig.\ref{fig:fey-c}),
  the light spectator quark in the $\overline{B}_{q}^{\ast}$ meson is
  absorbed by the recoiled $D_{q}$ ($M_{q}$) meson, and the exchanged
  gluons are space-like.
  In the annihilation topologies of Fig.\ref{fig:fey-a}, the exchanged
  gluons are time-like, which then split into the light quark-antiquark
  pair.


  The first two diagrams of Fig.\ref{fig:fey-t}, Fig.\ref{fig:fey-c},
  and Fig.\ref{fig:fey-a} are usually called the factorizable topologies.
  In the color-allowed (color-suppressed) factorizable emission topologies,
  the gluons are exchanged only between the initial $\overline{B}_{q}^{\ast}$
  and the recoil $D_{q}$ ($M_{q}$) meson pair, and the emission $M$ ($D^{0}$)
  meson could be completely parted from the $\overline{B}_{q}^{\ast}D_{q}$
  ($\overline{B}_{q}^{\ast}M_{q}$) system.
  In the factorizable annihilation topologies, the gluons are exchanged
  only between the final $DM$ meson pair, and the initial $\overline{B}_{q}^{\ast}$
  meson could be directly separated from the $DM$ meson pair.
  Hence, in the factorizable emission (annihilation) topologies, the integral
  of the wave functions for the emission (initial) mesons reduces to the
  corresponding decay constant.
  For the factorizable topologies, the decay amplitudes will have the relatively
  simple structures, and can be written as the product of the decay constants
  and the hadron transition form factors.
  With the pQCD approach, the form factors can be written as the convolution
  integral of the hard scattering amplitudes and the hadron DAs.

  The last two diagrams of Fig.\ref{fig:fey-t}, Fig.\ref{fig:fey-c}, and
  Fig.\ref{fig:fey-a} are usually called the nonfactorizable topologies.
  In the nonfactorizable topologies, the emission meson is entangled
  with the gluons that radiated from the spectator quark, and hence
  on meson can be separated clearly from the other mesons.
  Hence, the decay amplitudes for the nonfactorizable topologies
  have quite complicated structures, and the amplitude convolution
  integral involve the wave functions of all the participating mesons.
  The nonfactorizable emission topologies within the pQCD framework
  are also called the spectator scattering topologies with the QCDF
  approach.
  Especially for the color-suppressed emission topologies, the factorizable
  contributions are proportional to the small parameter $a_{2}$, hence,
  the nonfactorizable contributions, being proportional to the large
  Wilson coefficient $C_{1}$, should be significant.
  As widely recognized, the nonfactorizable contributions play
  an important role in clarifying or reducing some discrepancies
  between the theoretical results and the experimental data on the
  nonleptonic $B$ meson weak decays.

  Among the three possible types of Feynman diagrams (Fig.\ref{fig:fey-t},
  Fig.\ref{fig:fey-c}, and Fig.\ref{fig:fey-a}), only one or two of them
  will contribute to the specific $\overline{B}_{q}^{\ast}$ ${\to}$ $DM$ decays.
  The explicit amplitudes for the concrete $\overline{B}_{q}^{\ast}$ ${\to}$ $DP$, $DV$
  decays have been collected in the Appendixes \ref{amp-dp} and \ref{amp-dv}, and
  the building blocks in the Appendixes \ref{block-t}, \ref{block-c} and \ref{block-a}.
  According to the polarization relations between the initial and final vector
  mesons, the amplitudes for the $\overline{B}_{q}^{\ast}$ ${\to}$ $DV$ decays
  can generally be decomposed into the following structures
  \cite{prd66.054013,ijmpa31.1650146,npb911.890,prd95.036024},
   \begin{equation}
  {\cal A}(\overline{B}_{q}^{\ast}{\to}DV)\, =\,
  {\cal A}_{L}({\epsilon}_{B_{q}^{\ast}}^{\parallel},{\epsilon}_{V}^{\parallel})
 +{\cal A}_{N}({\epsilon}_{B_{q}^{\ast}}^{\perp}{\cdot}{\epsilon}_{V}^{\perp})
 +i\,{\cal A}_{T}\,{\varepsilon}_{{\mu}{\nu}{\alpha}{\beta}}\,
  {\epsilon}_{B_{q}^{\ast}}^{\mu}\,{\epsilon}_{V}^{\nu}\,
   p_{B_{q}^{\ast}}^{\alpha}\,p_{V}^{\beta}
   \label{eq:amp01}.
   \end{equation}
  which is conventionally written as the helicity amplitudes,
   \begin{equation}
   H_{0}\ =\ {\cal A}_{L}({\epsilon}_{B_{q}^{\ast}}^{\parallel},{\epsilon}_{V}^{\parallel})
   \label{eq:amp02},
   \end{equation}
   \begin{equation}
   H_{\parallel}\ =\ \sqrt{2}\,{\cal A}_{N}
   \label{eq:amp03},
   \end{equation}
   \begin{equation}
   H_{\perp}\ =\ \sqrt{2}\,m_{B_{q}^{\ast}}\,p\, {\cal A}_{T}
   \label{eq:amp04}.
   \end{equation}

  As is well known, it is commonly assumed that the $SU(3)$ symmetry breaking
  interactions mixes the isospin-singlet neutral members of the octet with
  the singlet states.
  The ideal mixing angle ${\theta}_{V}$ (with ${\sin}{\theta}_{V}$ $=$ $1/\sqrt{3}$)
  between the octet and the singlet states is almost true in practice for
  the physical ${\omega}$ and ${\phi}$ mesons, i.e.,
  the valence quark components are ${\omega}$ $=$ $(u\bar{u}+d\bar{d})/\sqrt{2}$
  and ${\phi}$ $=$ $s\bar{s}$.
  As for the mixing among the light pseudoscalar mesons, the notations known as
  the quark-flavor basis description \cite{prd58.114006} is adopted here, and for
  simplicity, the possible gluonium and charmonium compositions are neglected
  for the time being, i.e.,
   \begin{equation}
   \left(\begin{array}{c}
  {\eta} \\ {\eta}^{\prime}
   \end{array}\right) =
   \left(\begin{array}{cc}
  {\cos}{\theta}_{P} & -{\sin}{\theta}_{P} \\
  {\sin}{\theta}_{P} &  {\cos}{\theta}_{P}
   \end{array}\right)
   \left(\begin{array}{c}
  {\eta}_{q} \\ {\eta}_{s}
   \end{array}\right)
   \label{mixing01},
   \end{equation}
  where the flavor states ${\eta}_{q}$ $=$ $(u\bar{u}+d\bar{d})/{\sqrt{2}}$ and
  ${\eta}_{s}$ $=$ $s\bar{s}$. The mixing angle determined from experimental
  data is ${\theta}_{P}$ $=$ $(39.3{\pm}1.0)^{\circ}$ \cite{prd58.114006}.
  The mass relations between the physical states (${\eta}$
  and ${\eta}^{\prime}$) and the flavor states (${\eta}_{q}$
  and ${\eta}_{s}$) are
   \begin{eqnarray}
   m_{{\eta}_{q}}^{2}&=& \displaystyle
   m_{\eta}^{2}\,{\cos}^{2}{\theta}_{P}
  +m_{{\eta}^{\prime}}^{2}\,{\sin}^{2}{\theta}_{P}
  -\frac{\sqrt{2}\,f_{{\eta}_{s}}}{f_{{\eta}_{q}}}\,
  (m_{{\eta}^{\prime}}^{2}- m_{\eta}^{2})\,
  {\cos}{\theta}_{P}\,{\sin}{\theta}_{P}
   \label{ss12}, \\
   m_{{\eta}_{s}}^{2}&=& \displaystyle
   m_{\eta}^{2}\,{\sin}^{2}{\theta}_{P}
  +m_{{\eta}^{\prime}}^{2}\,{\cos}^{2}{\theta}_{P}
  -\frac{f_{{\eta}_{q}}}{\sqrt{2}\,f_{{\eta}_{s}}}
  (m_{{\eta}^{\prime}}^{2}- m_{\eta}^{2})\,
  {\cos}{\theta}_{P}\,{\sin}{\theta}_{P}
   \label{ss13},
   \end{eqnarray}
  where $f_{{\eta}_{q}}$ and $f_{{\eta}_{s}}$ are the decay constants.

  The amplitudes for the $\overline{B}_{q}^{\ast}$ ${\to}$ $D{\eta}$,
  $D{\eta}^{\prime}$ decays can be written as
   \begin{eqnarray}
  {\cal A}(\overline{B}_{q}^{\ast}{\to}D{\eta})
   &=&
   {\cos}{\theta}_{P}\,{\cal A}(\overline{B}_{q}^{\ast}{\to}D{\eta}_{q})
  -{\sin}{\theta}_{P}\,{\cal A}(\overline{B}_{q}^{\ast}{\to}D{\eta}_{s})
    \label{amp-bu-eta}, \\
   {\cal A}(\overline{B}_{q}^{\ast}{\to}D{\eta}^{\prime})
   &=&
   {\sin}{\theta}_{P}\,{\cal A}(\overline{B}_{q}^{\ast}{\to}D{\eta}_{q})
  +{\cos}{\theta}_{P}\,{\cal A}(\overline{B}_{q}^{\ast}{\to}D{\eta}_{s})
    \label{amp-bu-etap}.
    \end{eqnarray}

  \section{Numerical results and discussion}
  \label{sec03}
  In the rest frame of the $\overline{B}_{q}^{\ast}$ meson,
  the branching ratio is defined as
   \begin{equation}
  {\cal B}r(\overline{B}_{q}^{\ast}{\to}DV)\, =\, \frac{1}{24{\pi}}\,
   \frac{p}{m_{B^{\ast}_{q}}^{2}{\Gamma}_{B^{\ast}_{q}}}\, \Big\{
  {\vert}H_{0}{\vert}^{2}+{\vert}H_{\parallel}{\vert}^{2}
 +{\vert}H_{\perp}{\vert}^{2} \Big\}
   \label{br-dv},
   \end{equation}
   \begin{equation}
  {\cal B}r(\overline{B}_{q}^{\ast}{\to}DP)\, =\, \frac{1}{24{\pi}}\,
   \frac{p}{m_{B^{\ast}_{q}}^{2}{\Gamma}_{B^{\ast}_{q}}}\,
  {\vert}{\cal A}(\overline{B}_{q}^{\ast}{\to}DP){\vert}^{2}
   \label{br-dp},
   \end{equation}
  where ${\Gamma}_{B^{\ast}_{q}}$ is the full decay width
  of the $\overline{B}_{q}^{\ast}$ meson.

  Unfortunately, the experimental data on ${\Gamma}_{B^{\ast}_{q}}$
  are still unavailable until now.
  As is generally known, the electromagnetic radiation processes
  $\overline{B}^{\ast}_{q}$ ${\to}$ $\overline{B}_{q}{\gamma}$ dominate
  the $\overline{B}^{\ast}_{q}$ meson decays, and the mass
  differences between the $\overline{B}^{\ast}_{q}$ and $\overline{B}_{q}$
  mesons are very small, $m_{B_{q}^{\ast}}$ $-$ $m_{B_{q}}$ ${\lesssim}$
  50 MeV \cite{pdg}, which results in the fact that the photons from
  the $\overline{B}^{\ast}_{q}$ ${\to}$ $\overline{B}_{q}{\gamma}$ process
  are too soft to be easily identified by the detectors at the existing
  experiments.
  A good approximation for the decay width is ${\Gamma}_{B_{q}^{\ast}}$
  ${\approx}$ ${\Gamma}(\overline{B}_{q}^{\ast}{\to}\overline{B}_{q}{\gamma})$.
  Theoretically, there is the close relation between the partial decay width
  for the $\overline{B}_{q}^{\ast}$ ${\to}$ $\overline{B}_{q}{\gamma}$ decay
  and the magnetic dipole (M1) moment of the $\overline{B}_{q}^{\ast}$ meson
  \cite{epja52.90}, i.e.,
   \begin{equation}
  {\Gamma}(\overline{B}_{q}^{\ast}{\to}\overline{B}_{q}{\gamma})\, =\,
   \frac{4}{3}\,{\alpha}_{\rm em}\, k_{\gamma}^{3}\, {\mu}^{2}_{h}
   \label{m1-width},
   \end{equation}
  where ${\alpha}_{\rm em}$ is the fine structure constant;
  $k_{\gamma}$ $=$ $(m_{B_{q}^{\ast}}^{2}-m_{B_{q}}^{2})/2m_{B_{q}^{\ast}}$
  is the photon momentum in the rest frame of the $\overline{B}_{q}^{\ast}$
  meson; ${\mu}_{h}$ is the M1 moment of the $\overline{B}_{q}^{\ast}$ meson.
  There are a large number of theoretical predictions on the partial decay
  width ${\Gamma}(\overline{B}_{q}^{\ast}{\to}\overline{B}_{q}{\gamma})$.
  Many of these have been collected into Table 7 of Ref.\cite{jhep1404.177}
  and Tables 3 and 4 of Ref.\cite{epja52.90}.
  However, there are big differences among these estimations with
  various models, due to our inaccurate information about the M1
  moments of mesons.
  In principle, the M1 moment of a hadron should be the sum of the
  M1 moments of its constituent quarks.
  As is well known, for an elementary particle, the M1 moment is
  proportional to the charge and inversely proportional to the mass.
  Hence, the M1 moment of the heavy-light $\overline{B}_{q}^{\ast}$
  meson should be mainly affected by the M1 moment of the light quark
  rather than the bottom quark.
  With the M1 moments of the light $u$, $d$ and $s$ quarks in the
  terms of the nuclear magnetons ${\mu}_{N}$, i.e.,
  ${\mu}_{u}$ ${\simeq}$ $1.85\,{\mu}_{N}$,
  ${\mu}_{d}$ ${\simeq}$ $-0.97\,{\mu}_{N}$, and
  ${\mu}_{s}$ ${\simeq}$ $-0.61\,{\mu}_{N}$ \cite{uds},
  it is expected to have the relations
  ${\Gamma}(\overline{B}_{u}^{\ast}{\to}\overline{B}_{u}{\gamma})$ $>$
  ${\Gamma}(\overline{B}_{d}^{\ast}{\to}\overline{B}_{d}{\gamma})$ $>$
  ${\Gamma}(\overline{B}_{s}^{\ast}{\to}\overline{B}_{s}{\gamma})$, and therefore the relations
  ${\Gamma}_{B_{u}^{\ast}}$ $>$ ${\Gamma}_{B_{d}^{\ast}}$
  $>$ ${\Gamma}_{B_{s}^{\ast}}$.
  It is far beyond the scope of this paper to elaborate more on the
  details of the decay width ${\Gamma}_{B_{q}^{\ast}}$.
  In our calculation, in order to give a quantitative estimation of
  the branching ratios for the $\overline{B}_{q}^{\ast}$ ${\to}$ $DM$
  decays, we will use the following values of the decay widths,
   \begin{equation}
  {\Gamma}_{B_{u}^{\ast}}\ {\sim}\
  {\Gamma}(\overline{B}_{u}^{\ast}{\to}\overline{B}_{u}{\gamma})\ {\sim}\
   450\,\text{eV}
   \label{m1-width-u},
   \end{equation}
   \begin{equation}
  {\Gamma}_{B_{d}^{\ast}}\ {\sim}\
  {\Gamma}(\overline{B}_{d}^{\ast}{\to}\overline{B}_{d}{\gamma})\ {\sim}\
   150\,\text{eV}
   \label{m1-width-d},
   \end{equation}
   \begin{equation}
  {\Gamma}_{B_{s}^{\ast}}\ {\sim}\
  {\Gamma}(\overline{B}_{s}^{\ast}{\to}\overline{B}_{s}{\gamma})\ {\sim}\
   100\,\text{eV}
   \label{m1-width-s},
   \end{equation}
  which is basically consistent with the recent results
  of Ref.\cite{epja52.90}.

  \begin{table}[ht]
  \caption{The numerical values of input parameters.}
  \label{tab:input}
  \begin{ruledtabular}
  \begin{tabular}{lll}
    CKM parameter
  & $A$ $=$ $0.811{\pm}0.026$ \cite{pdg},
  & ${\lambda}$ $=$ $0.22506{\pm}0.00050$ \cite{pdg}, \\ \hline
    mass of the particles
  & $m_{{\pi}^{\pm}}$ $=$ $139.57$ MeV \cite{pdg},
  & $m_{K^{\pm}}$ $=$ $493.677{\pm}0.016$ MeV \cite{pdg}, \\
    $m_{b}$ $=$ $4.78{\pm}0.06$ GeV \cite{pdg},
  & $m_{{\pi}^{0}}$ $=$ $134.98$ MeV \cite{pdg},
  & $m_{K^{0}}$ $=$ $497.611{\pm}0.013$ MeV \cite{pdg}, \\
    $m_{c}$ $=$ $1.67{\pm}0.07$ GeV \cite{pdg},
  & $m_{{\eta}^{\prime}}$ $=$ $957.78{\pm}0.06$ MeV \cite{pdg},
  & $m_{\eta}$ $=$ $547.862{\pm}0.017$ MeV \cite{pdg}, \\
    $m_{s}$ ${\simeq}$ $0.51$ GeV \cite{uds},
  & $m_{\rho}$ $=$ $775.26{\pm}0.25$ MeV \cite{pdg},
  & $m_{K^{{\ast}0}}$ $=$ $895.81{\pm}0.19$ MeV \cite{pdg}, \\
    $m_{u,d}$ ${\simeq}$ $0.31$ GeV \cite{uds},
  & $m_{\omega}$ $=$ $782.62{\pm}0.12$ MeV \cite{pdg},
  & $m_{K^{{\ast}{\pm}}}$ $=$ $891.66{\pm}0.26$ MeV \cite{pdg}, \\
    $m_{B_{s}^{\ast}}$ $=$ $5415.4^{+1.8}_{-1.5}$ MeV \cite{pdg},
  & $m_{B_{u,d}^{\ast}}$ $=$ $5324.65{\pm}0.25$ MeV \cite{pdg},
  & $m_{\phi}$ $=$ $1019.461{\pm}0.019$ MeV \cite{pdg}, \\
    $m_{D_{s}}$ $=$ $1968.27{\pm}0.10$ MeV \cite{pdg},
  & $m_{D_{d}}$ $=$ $1869.58{\pm}0.09$ MeV \cite{pdg},
  & $m_{D_{u}}$ $=$ $1864.83{\pm}0.05$ MeV \cite{pdg}, \\ \hline
    decay constant
  & $f_{\pi}$ $=$ $130.2{\pm}1.7$ MeV \cite{pdg},
  & $f_{K}$ $=$ $155.6{\pm}0.4$ MeV \cite{pdg}, \\
    $f_{{\eta}_{q}}$ $=$ $(1.07{\pm}0.02)\,f_{\pi}$ \cite{prd58.114006},
  & $f_{K^{\ast}}$ $=$ $220{\pm}5$ MeV \cite{jhep0703.069},
  & $f_{K^{\ast}}^{T}$ $=$ $185{\pm}10$ MeV \cite{jhep0703.069}, \\
    $f_{{\eta}_{s}}$ $=$ $(1.34{\pm}0.06)\,f_{\pi}$ \cite{prd58.114006},
  & $f_{\rho}$ $=$ $216{\pm}3$ MeV \cite{jhep0703.069},
  & $f_{\rho}^{T}$ $=$ $165{\pm}9$ MeV \cite{jhep0703.069}, \\
    $f_{D_{s}}$ $=$ $249.0{\pm}1.2$ MeV \cite{pdg},
  & $f_{\omega}$ $=$ $187{\pm}5$ MeV \cite{jhep0703.069},
  & $f_{\omega}^{T}$ $=$ $151{\pm}9$ MeV \cite{jhep0703.069}, \\
    $f_{D_{u,d}}$ $=$ $211.9{\pm}1.1$ MeV \cite{pdg},
  & $f_{\phi}$ $=$ $215{\pm}5$ MeV \cite{jhep0703.069},
  & $f_{\phi}^{T}$ $=$ $186{\pm}9$ MeV \cite{jhep0703.069}, \\
    $f_{B_{s}^{\ast}}$ $=$ $213{\pm}7$ MeV \cite{prd91.114509},
  & $f_{B_{u,d}^{\ast}}$ $=$ $175{\pm}6$ MeV \cite{prd91.114509},
  & \\ \hline
    Gegenbauer moment
  & $a_{2}^{{\pi},{\eta}_{q,s}}$ $=$ $0.25{\pm}0.15$ \cite{jhep0605.004},
  & $a_{2}^{{\parallel},{\rho},{\omega}}$ $=$ $0.15{\pm}0.07$ \cite{jhep0703.069}, \\
    $a_{1}^{K}$ $=$ $-0.06{\pm}0.03$ \cite{jhep0605.004},
  & $a_{2}^{K}$ $=$ $0.25{\pm}0.15$ \cite{jhep0605.004},
  & $a_{2}^{{\perp},{\rho},{\omega}}$ $=$ $0.14{\pm}0.06$ \cite{jhep0703.069}, \\
    $a_{1}^{{\parallel},K^{\ast}}$ $=$ $-0.03{\pm}0.02$ \cite{jhep0703.069},
  & $a_{2}^{{\parallel},K^{\ast}}$ $=$ $0.11{\pm}0.09$ \cite{jhep0703.069},
  & $a_{2}^{{\parallel},{\phi}}$ $=$ $0.18{\pm}0.08$ \cite{jhep0703.069}, \\
    $a_{1}^{{\perp},K^{\ast}}$ $=$ $-0.04{\pm}0.03$ \cite{jhep0703.069},
  & $a_{2}^{{\perp},K^{\ast}}$ $=$ $0.10{\pm}0.08$ \cite{jhep0703.069},
  & $a_{2}^{{\perp},{\phi}}$ $=$ $0.14{\pm}0.07$ \cite{jhep0703.069}.
  \end{tabular}
  \end{ruledtabular}
  \end{table}

  \begin{table}[ht]
  \caption{The branching ratios for the $\overline{B}_{q}^{\ast}$
  ${\to}$ $DP$ decays with the different DA scenarios,
  where the theoretical uncertainties come from the scale $(1{\pm}0.1)t_{i}$,
  the mass $m_{c}$ and $m_{b}$, and the hadronic parameters (including
  the decay constants, Gegenbauer moments, and so on).}
  \label{tab:br-dp}
  \begin{ruledtabular}
  \begin{tabular}{lccccc}
    decay mode & class & unit & I & II & III \\ \hline
    $B_{u}^{{\ast}-}$ ${\to}$             $D_{u}^{0}{\pi}^{-}$
  & T-I  & $10^{-10}$
  & $ 6.61^{+ 2.12+ 0.07+ 0.75}_{- 0.92- 0.79- 0.69}$
  & $ 1.25^{+ 0.25+ 0.11+ 0.16}_{- 0.13- 0.13- 0.15}$
  & $ 0.56^{+ 0.13+ 0.09+ 0.08}_{- 0.07- 0.09- 0.07}$
  \\
    $B_{u}^{{\ast}-}$ ${\to}$             $D_{u}^{0}K^{-}$
  & T-II & $10^{-11}$
  & $ 5.38^{+ 1.97+ 0.05+ 0.55}_{- 0.85- 0.63- 0.51}$
  & $ 0.95^{+ 0.20+ 0.09+ 0.11}_{- 0.10- 0.10- 0.10}$
  & $ 0.42^{+ 0.10+ 0.07+ 0.05}_{- 0.05- 0.07- 0.05}$
  \\
    $\overline{B}_{d}^{{\ast}0}$ ${\to}$  $D_{d}^{+}{\pi}^{-}$
  & T-I  & $10^{ -9}$
  & $ 2.22^{+ 0.52+ 0.02+ 0.27}_{- 0.20- 0.23- 0.24}$
  & $ 0.51^{+ 0.08+ 0.04+ 0.07}_{- 0.03- 0.05- 0.06}$
  & $ 0.28^{+ 0.04+ 0.03+ 0.04}_{- 0.02- 0.04- 0.03}$
  \\
    $\overline{B}_{d}^{{\ast}0}$ ${\to}$  $D_{d}^{+}K^{-}$
  & T-II & $10^{-10}$
  & $ 1.69^{+ 0.38+ 0.01+ 0.15}_{- 0.15- 0.18- 0.14}$
  & $ 0.38^{+ 0.06+ 0.03+ 0.03}_{- 0.02- 0.03- 0.03}$
  & $ 0.20^{+ 0.03+ 0.03+ 0.02}_{- 0.01- 0.03- 0.02}$
  \\
    $\overline{B}_{d}^{{\ast}0}$ ${\to}$  $D_{u}^{0}{\pi}^{0}$
  & C-I  & $10^{-12}$
  & $ 4.04^{+ 3.19+ 0.38+ 1.32}_{- 2.47- 0.39- 1.05}$
  & $ 6.11^{+ 0.81+ 0.32+ 1.72}_{- 0.44- 0.29- 1.28}$
  & $ 7.21^{+ 1.02+ 0.13+ 1.66}_{- 0.88- 0.09- 1.35}$
  \\
    $\overline{B}_{d}^{{\ast}0}$ ${\to}$  $D_{u}^{0}{\eta}$
  & C-I  & $10^{-12}$
  & $ 3.48^{+ 2.78+ 0.21+ 0.80}_{- 2.29- 0.11- 1.28}$
  & $ 4.49^{+ 0.63+ 0.36+ 0.98}_{- 0.43- 0.25- 0.84}$
  & $ 3.60^{+ 0.73+ 0.22+ 0.68}_{- 0.71- 0.16- 0.60}$
  \\
    $\overline{B}_{d}^{{\ast}0}$ ${\to}$  $D_{u}^{0}{\eta}^{\prime}$
  & C-I  & $10^{-12}$
  & $ 2.26^{+ 1.80+ 0.13+ 0.67}_{- 1.49- 0.07- 0.72}$
  & $ 2.91^{+ 0.41+ 0.23+ 0.68}_{- 0.28- 0.16- 0.56}$
  & $ 2.33^{+ 0.47+ 0.14+ 0.49}_{- 0.46- 0.10- 0.41}$
  \\
    $\overline{B}_{d}^{{\ast}0}$ ${\to}$  $D_{u}^{0}\overline{K}^{0}$
  & C-II & $10^{-13}$
  & $ 2.19^{+ 4.55+ 0.72+ 2.30}_{- 0.44- 0.51- 1.40}$
  & $ 9.66^{+ 1.59+ 0.72+ 2.86}_{- 1.16- 0.50- 2.14}$
  & $ 9.54^{+ 1.94+ 0.39+ 1.96}_{- 1.92- 0.27- 1.72}$
  \\
    $\overline{B}_{d}^{{\ast}0}$ ${\to}$  $D_{s}^{+}K^{-}$
  & A-I  & $10^{-12}$
  & $ 0.64^{+ 0.05+ 0.17+ 1.09}_{- 0.04- 0.16- 0.42}$
  & $ 1.30^{+ 0.06+ 0.10+ 0.89}_{- 0.01- 0.09- 0.40}$
  & $ 0.40^{+ 0.05+ 0.03+ 0.20}_{- 0.01- 0.03- 0.08}$
  \\
    $\overline{B}_{s}^{{\ast}0}$ ${\to}$  $D_{s}^{+}{\pi}^{-}$
  & T-I  & $10^{ -9}$
  & $ 5.68^{+ 1.17+ 0.09+ 0.60}_{- 0.46- 0.50- 0.56}$
  & $ 1.48^{+ 0.21+ 0.07+ 0.16}_{- 0.09- 0.12- 0.15}$
  & $ 0.81^{+ 0.11+ 0.07+ 0.09}_{- 0.05- 0.09- 0.08}$
  \\
    $\overline{B}_{s}^{{\ast}0}$ ${\to}$  $D_{s}^{+}K^{-}$
  & T-II & $10^{-10}$
  & $ 4.30^{+ 0.89+ 0.09+ 0.40}_{- 0.35- 0.37- 0.38}$
  & $ 1.17^{+ 0.17+ 0.06+ 0.12}_{- 0.07- 0.09- 0.11}$
  & $ 0.64^{+ 0.09+ 0.06+ 0.06}_{- 0.04- 0.07- 0.06}$
  \\
    $\overline{B}_{s}^{{\ast}0}$ ${\to}$  $D_{d}^{+}{\pi}^{-}$
  & A-II & $10^{-14}$
  & $ 0.16^{+ 0.17+ 0.75+ 2.36}_{- 0.02- 0.15- 0.08}$
  & $ 1.71^{+ 0.31+ 0.48+ 2.44}_{- 0.05- 0.46- 0.40}$
  & $ 1.80^{+ 0.58+ 0.10+ 2.08}_{- 0.24- 0.10- 0.58}$
  \\
    $\overline{B}_{s}^{{\ast}0}$ ${\to}$  $D_{u}^{0}{\pi}^{0}$
  & A-II & $10^{-14}$
  & $ 0.08^{+ 0.09+ 0.38+ 1.19}_{- 0.01- 0.08- 0.04}$
  & $ 0.86^{+ 0.16+ 0.24+ 1.23}_{- 0.03- 0.23- 0.20}$
  & $ 0.90^{+ 0.29+ 0.05+ 1.05}_{- 0.12- 0.05- 0.29}$
  \\
    $\overline{B}_{s}^{{\ast}0}$ ${\to}$  $D_{u}^{0}{\eta}$
  & C-II & $10^{-12}$
  & $ 0.20^{+ 0.38+ 0.06+ 0.15}_{- 0.02- 0.06- 0.09}$
  & $ 0.90^{+ 0.16+ 0.05+ 0.26}_{- 0.12- 0.07- 0.20}$
  & $ 1.01^{+ 0.21+ 0.01+ 0.27}_{- 0.21- 0.03- 0.22}$
  \\
    $\overline{B}_{s}^{{\ast}0}$ ${\to}$  $D_{u}^{0}{\eta}^{\prime}$
  & C-II & $10^{-12}$
  & $ 0.36^{+ 0.72+ 0.10+ 0.21}_{- 0.08- 0.08- 0.16}$
  & $ 1.38^{+ 0.24+ 0.08+ 0.31}_{- 0.19- 0.09- 0.31}$
  & $ 1.22^{+ 0.30+ 0.02+ 0.23}_{- 0.30- 0.06- 0.26}$
  \\
    $\overline{B}_{s}^{{\ast}0}$ ${\to}$  $D_{u}^{0}K^{0}$
  & C-I  & $10^{-11}$
  & $ 3.31^{+ 2.07+ 0.12+ 1.09}_{- 1.88- 0.24- 0.77}$
  & $ 4.31^{+ 0.51+ 0.16+ 1.22}_{- 0.28- 0.25- 0.93}$
  & $ 4.21^{+ 0.61+ 0.02+ 0.94}_{- 0.53- 0.16- 0.80}$
  \end{tabular}
  \end{ruledtabular}
  \end{table}

  \begin{table}[ht]
  \caption{The branching ratios for the $\overline{B}_{q}^{\ast}$
  ${\to}$ $DV$ decays with the different DA scenarios,
  where the theoretical uncertainties arise from the scale $(1{\pm}0.1)t_{i}$,
  the mass $m_{c}$ and $m_{b}$, and the hadronic parameters (including
  the decay constants, Gegenbauer moments,and so on).}
  \label{tab:br-dv}
  \begin{ruledtabular}
  \begin{tabular}{lccccc}
    decay mode & class & unit & Scenario I & Scenario II & Scenario III \\ \hline
    $B_{u}^{{\ast}-}$ ${\to}$             $D_{u}^{0}{\rho}^{-}$
  & T-I  & $10^{ -9}$
  & $ 2.02^{+ 0.55+ 0.02+ 0.23}_{- 0.23- 0.22- 0.21}$
  & $ 0.43^{+ 0.08+ 0.04+ 0.05}_{- 0.04- 0.04- 0.05}$
  & $ 0.19^{+ 0.04+ 0.03+ 0.02}_{- 0.02- 0.03- 0.02}$
  \\
    $B_{u}^{{\ast}-}$ ${\to}$             $D_{u}^{0}K^{{\ast}-}$
  & T-II & $10^{-10}$
  & $ 1.14^{+ 0.32+ 0.01+ 0.16}_{- 0.13- 0.13- 0.14}$
  & $ 0.25^{+ 0.05+ 0.02+ 0.04}_{- 0.02- 0.03- 0.03}$
  & $ 0.11^{+ 0.02+ 0.02+ 0.02}_{- 0.01- 0.02- 0.01}$
  \\
    $\overline{B}_{d}^{{\ast}0}$ ${\to}$  $D_{d}^{+}{\rho}^{-}$
  & T-I  & $10^{ -9}$
  & $ 6.80^{+ 1.55+ 0.03+ 0.79}_{- 0.60- 0.66- 0.72}$
  & $ 1.72^{+ 0.26+ 0.11+ 0.21}_{- 0.11- 0.15- 0.19}$
  & $ 0.95^{+ 0.13+ 0.11+ 0.11}_{- 0.06- 0.11- 0.10}$
  \\
    $\overline{B}_{d}^{{\ast}0}$ ${\to}$  $D_{d}^{+}K^{{\ast}-}$
  & T-II & $10^{-10}$
  & $ 3.87^{+ 0.87+ 0.01+ 0.51}_{- 0.33- 0.36- 0.46}$
  & $ 0.99^{+ 0.15+ 0.07+ 0.13}_{- 0.06- 0.09- 0.12}$
  & $ 0.53^{+ 0.07+ 0.07+ 0.07}_{- 0.03- 0.07- 0.07}$
  \\
    $\overline{B}_{d}^{{\ast}0}$ ${\to}$  $D_{u}^{0}{\rho}^{0}$
  & C-I  & $10^{-11}$
  & $ 2.55^{+ 1.26+ 0.15+ 0.57}_{- 1.06- 0.09- 0.47}$
  & $ 4.60^{+ 0.61+ 0.15+ 0.84}_{- 0.44- 0.09- 0.72}$
  & $ 5.95^{+ 0.79+ 0.23+ 1.14}_{- 0.76- 0.19- 0.98}$
  \\
    $\overline{B}_{d}^{{\ast}0}$ ${\to}$  $D_{u}^{0}{\omega}$
  & C-I  & $10^{-11}$
  & $ 2.43^{+ 1.00+ 0.17+ 0.75}_{- 0.79- 0.11- 0.60}$
  & $ 4.32^{+ 0.51+ 0.21+ 1.11}_{- 0.35- 0.15- 0.91}$
  & $ 4.71^{+ 0.64+ 0.16+ 1.14}_{- 0.62- 0.14- 0.94}$
  \\
    $\overline{B}_{d}^{{\ast}0}$ ${\to}$  $D_{u}^{0}\overline{K}^{{\ast}0}$
  & C-II & $10^{-12}$
  & $ 3.58^{+ 1.77+ 0.22+ 1.14}_{- 1.52- 0.16- 0.89}$
  & $ 6.26^{+ 0.90+ 0.32+ 1.70}_{- 0.70- 0.25- 1.40}$
  & $ 7.83^{+ 1.16+ 0.28+ 2.17}_{- 1.16- 0.29- 1.76}$
  \\
    $\overline{B}_{d}^{{\ast}0}$ ${\to}$  $D_{s}^{+}K^{{\ast}-}$
  & A-I  & $10^{-12}$
  & $ 5.55^{+ 0.94+ 0.50+ 2.57}_{- 0.46- 0.48- 1.68}$
  & $ 6.82^{+ 0.58+ 0.54+ 2.09}_{- 0.13- 0.49- 1.43}$
  & $ 3.98^{+ 0.54+ 0.10+ 1.19}_{- 0.18- 0.08- 0.78}$
  \\
    $\overline{B}_{s}^{{\ast}0}$ ${\to}$  $D_{s}^{+}{\rho}^{-}$
  & T-I  & $10^{ -8}$
  & $ 1.72^{+ 0.35+ 0.03+ 0.19}_{- 0.14- 0.13- 0.17}$
  & $ 0.50^{+ 0.07+ 0.03+ 0.05}_{- 0.03- 0.04- 0.05}$
  & $ 0.27^{+ 0.04+ 0.02+ 0.03}_{- 0.02- 0.03- 0.03}$
  \\
    $\overline{B}_{s}^{{\ast}0}$ ${\to}$  $D_{s}^{+}K^{{\ast}-}$
  & T-II & $10^{ -9}$
  & $ 1.00^{+ 0.21+ 0.02+ 0.13}_{- 0.08- 0.09- 0.12}$
  & $ 0.31^{+ 0.05+ 0.02+ 0.04}_{- 0.02- 0.03- 0.04}$
  & $ 0.17^{+ 0.02+ 0.01+ 0.02}_{- 0.01- 0.02- 0.02}$
  \\
    $\overline{B}_{s}^{{\ast}0}$ ${\to}$  $D_{d}^{+}{\rho}^{-}$
  & A-II & $10^{-13}$
  & $ 2.39^{+ 0.79+ 0.47+ 0.90}_{- 0.31- 0.20- 0.53}$
  & $ 6.23^{+ 0.26+ 0.86+ 1.27}_{- 0.00- 0.72- 0.91}$
  & $ 2.88^{+ 0.31+ 0.13+ 0.50}_{- 0.04- 0.11- 0.36}$
  \\
    $\overline{B}_{s}^{{\ast}0}$ ${\to}$  $D_{u}^{0}{\rho}^{0}$
  & A-II & $10^{-13}$
  & $ 1.19^{+ 0.40+ 0.24+ 0.45}_{- 0.15- 0.10- 0.27}$
  & $ 3.12^{+ 0.13+ 0.43+ 0.63}_{- 0.00- 0.36- 0.45}$
  & $ 1.44^{+ 0.15+ 0.06+ 0.25}_{- 0.02- 0.05- 0.18}$
  \\
    $\overline{B}_{s}^{{\ast}0}$ ${\to}$  $D_{u}^{0}{\omega}$
  & A-II & $10^{-13}$
  & $ 0.97^{+ 0.33+ 0.19+ 0.37}_{- 0.13- 0.08- 0.22}$
  & $ 2.42^{+ 0.11+ 0.34+ 0.55}_{- 0.00- 0.28- 0.40}$
  & $ 1.12^{+ 0.12+ 0.05+ 0.23}_{- 0.02- 0.04- 0.17}$
  \\
    $\overline{B}_{s}^{{\ast}0}$ ${\to}$  $D_{u}^{0}{\phi}$
  & C-II & $10^{-11}$
  & $ 0.79^{+ 0.32+ 0.03+ 0.18}_{- 0.30- 0.04- 0.15}$
  & $ 1.20^{+ 0.17+ 0.04+ 0.26}_{- 0.13- 0.04- 0.22}$
  & $ 1.50^{+ 0.22+ 0.02+ 0.31}_{- 0.21- 0.04- 0.27}$
  \\
    $\overline{B}_{s}^{{\ast}0}$ ${\to}$  $D_{u}^{0}K^{{\ast}0}$
  & C-I  & $10^{-10}$
  & $ 1.69^{+ 0.67+ 0.07+ 0.51}_{- 0.61- 0.09- 0.41}$
  & $ 2.58^{+ 0.32+ 0.07+ 0.70}_{- 0.23- 0.09- 0.57}$
  & $ 3.19^{+ 0.42+ 0.05+ 0.87}_{- 0.40- 0.11- 0.71}$
  \end{tabular}
  \end{ruledtabular}
  \end{table}

  The numerical values of other input parameters are collected in
  Table \ref{tab:input}, where their central values will be fixed
  as the default inputs unless otherwise specified.
  In addition, in order to investigate the effects from different
  DA models, we explore three scenarios,
  \begin{itemize}
  \item Scenario I:
   Eqs.(\ref{da-bqlv}-\ref{da-cqp})
   for the DAs of ${\phi}_{B^{\ast}}^{v,t,V,T}$ and ${\phi}_{D_{q}}^{a,p}$;
  \item Scenario II:
  ${\phi}_{B^{\ast}}^{v,t,V,T}$ $=$ Eq.(\ref{da-bqlv}),
  and ${\phi}_{D_{q}}^{a,p}$ $=$ Eq.(\ref{da-cqa});
  \item Scenario III:
  ${\phi}_{B^{\ast}}^{v,t,V,T}$ $=$ Eq.(\ref{da-bqlv}),
  and ${\phi}_{D_{q}}^{a,p}$ $=$ Eq.(\ref{wave-d-xb}).
  \end{itemize}

  Our numerical results on the branching ratios
  are presented in Tables \ref{tab:br-dp}
  and \ref{tab:br-dv}, where the uncertainties come from the
  typical scale $(1{\pm}0.1)t_{i}$, the mass $m_{c}$ and $m_{b}$,
  and the hadronic parameters (including the decay constants,
  Gegenbauer moments, and so on), respectively.
  The following are some comments.

  (1)
  Generally, the $\overline{B}_{q}^{\ast}$ ${\to}$ $DP$ decay modes
  could be divided into three categories, i.e. the ``T'', ``C'', and ``A''
  types are dominated by contributions from the color-allowed emission
  topologies of Fig.\ref{fig:fey-t}, the color-suppressed emission
  topologies of Fig.\ref{fig:fey-c}, and the pure annihilation topologies
  of Fig.\ref{fig:fey-a}, respectively.
  And each category could be further divided into two classes,
  i.e., the decay amplitudes of the classes ``I'' and ``II'' are
  proportional to the CKM factors of $V_{cb}\,V_{ud}^{\ast}$ ${\sim}$
  $A{\lambda}^{2}$ and $V_{cb}\,V_{us}^{\ast}$ ${\sim}$
  $A{\lambda}^{3}$, respectively.
  There are many hierarchical relations among the branching ratios,
  such as,
   \begin{equation}
  {\cal B}r(\text{class T-I}) >
  {\cal B}r(\text{class C-I}) >
  {\cal B}r(\text{class A-I})
   \label{r-01-01},
   \end{equation}
   \begin{equation}
  {\cal B}r(\text{class T-II}) >
  {\cal B}r(\text{class C-II}) >
  {\cal B}r(\text{class A-II})
   \label{r-01-02},
   \end{equation}
   \begin{equation}
  {\cal B}r(\text{class X-I}) >
  {\cal B}r(\text{class X-II}), \quad
   \text{for X\,=\,T,\,C,\,A}
   \label{r-01-03}.
   \end{equation}
  These categories and relations also happen to hold true
  for the $\overline{B}_{q}^{\ast}$ ${\to}$ $DV$ decays.

  For the ``T'' and ``C'' types of the $\overline{B}_{q}^{\ast}$
  ${\to}$ $DM$ decays, the annihilation contributions have a
  negligible impact on the branching ratios, and they are
  strongly suppressed relative to the emission contributions,
  as is stated by the QCDF approach \cite{npb591.313}.

  For the ``T'' types of the $\overline{B}_{q}^{\ast}$ ${\to}$
  $DM$ decays, the factorizable contributions from the emission
  topologies to the branching ratios are dominant over other
  contributions.
  However, for the ``C'' types of the $\overline{B}_{q}^{\ast}$ ${\to}$
  $DM$ decays, the nonfactorizable contributions to the branching
  ratios become very important, and sometimes even dominant.

  (2)
  With the law of conservation of angular momentum,
  three partial wave amplitudes, including the $s$-, $p$-, and $d$-wave
  amplitudes, all will contribute to the $\overline{B}_{q}^{\ast}$
  ${\to}$ $DV$ decays, while only the $p$-wave amplitude will
  contribute to the $\overline{B}_{q}^{\ast}$ ${\to}$ $DP$ decays.
  Besides, the branching ratios are proportional to the squares
  of the decay constants with the pQCD approach.
  With the magnitude relations between the decay constants
  $f_{V}$ $>$ $f_{P}$, one should expect to have the general
  relation of the branching ratios,
   \begin{equation}
  {\cal B}r(\overline{B}_{q}^{\ast}{\to}DV) >
  {\cal B}r(\overline{B}_{q}^{\ast}{\to}DP)
   \label{r-02-01},
   \end{equation}
  for the final vector $V$ and pseudoscalar $P$ mesons
  carrying the same flavor, azimuthal and magnetic isospin
  quantum numbers.
  And due to the relations between the decay constants
  $f_{B_{s}^{\ast}}$ $>$ $f_{B_{u,d}^{\ast}}$ and
  $f_{D_{s}}$ $>$ $f_{D_{u,d}}$,
  and the relations between the decay widths
  ${\Gamma}_{B_{s}^{\ast}}$ $<$ ${\Gamma}_{B_{u,d}^{\ast}}$,
  the color-allowed $\overline{B}_{s}^{{\ast}0}$ ${\to}$
  $D_{s}^{+}{\rho}^{-}$ decay has a relatively large
  branching ratio.

  Furthermore, our study results show that for the ``T''
  types of the $\overline{B}_{q}^{\ast}$ ${\to}$ $DV$ decays,
  the contributions of the longitudinal polarization
  part are dominant. Take the $\overline{B}_{s}^{{\ast}0}$
  ${\to}$ $D_{s}^{+}{\rho}^{-}$ decay for example,
  the longitudinal polarization fraction $f_{0}$ ${\equiv}$
  $\frac{{\vert}H_{0}{\vert}^{2}}{{\vert}H_{0}{\vert}^{2}
  +{\vert}H_{\parallel}{\vert}^{2}+{\vert}H_{\perp}{\vert}^{2}}$
  ${\approx}$ $90\%$ ($85\%$),
  the parallel polarization fraction $f_{\parallel}$ ${\equiv}$
  $\frac{{\vert}H_{0}{\vert}^{2}}{{\vert}H_{\parallel}{\vert}^{2}
  +{\vert}H_{\parallel}{\vert}^{2}+{\vert}H_{\perp}{\vert}^{2}}$
  ${\approx}$ $9\%$ ($12\%$),
  and the perpendicular polarization fraction $f_{\perp}$ ${\equiv}$
  $\frac{{\vert}H_{0}{\vert}^{2}}{{\vert}H_{\perp}{\vert}^{2}
  +{\vert}H_{\parallel}{\vert}^{2}+{\vert}H_{\perp}{\vert}^{2}}$
  ${\approx}$ $1\%$ ($3\%$) with the DA scenarios I (II and III),
  which generally agree with those obtained by the QCDF approach \cite{epjc76.523}.

  (3)
  As is well known, the theoretical results depend on the values
  of the input parameters.
  From the numbers in Tables \ref{tab:br-dp} and \ref{tab:br-dv},
  it is clearly seen that the main uncertainty is due to the
  limited knowledge of the hadron DAs,
  for example, the large discrepancy among the different DA scenarios.
  Besides the theoretical uncertainties listed in Tables \ref{tab:br-dp}
  and \ref{tab:br-dv}, the CKM parameters will bring some 6\%
  uncertainties.
  With a different value of the decay width ${\Gamma}_{B_{q}^{\ast}}$,
  the branching ratios in Tables \ref{tab:br-dp} and \ref{tab:br-dv}
  should be multiplied by the factors of
  ${450\,{\rm eV}}/{{\Gamma}_{B_{u}^{\ast}}}$,
  ${150\,{\rm eV}}/{{\Gamma}_{B_{d}^{\ast}}}$,
  ${100\,{\rm eV}}/{{\Gamma}_{B_{s}^{\ast}}}$
  for the $B_{u}^{\ast}$, $B_{d}^{\ast}$, $B_{s}^{\ast}$
  weak decays, respectively.
  To reduce the theoretical uncertainties, one of the
  commonly used methods is to exploit the rate of the
  branching ratios, such as,
   \begin{equation}
   \frac{{\cal B}r(\overline{B}_{u}^{{\ast}-}{\to}D_{u}^{0}{\pi}^{-})}
        {{\cal B}r(\overline{B}_{u}^{{\ast}-}{\to}D_{u}^{0}K^{-})}
   \ {\approx}\ \frac{f_{\pi}^{2}}{{\lambda}^{2}\,f_{K}^{2}}
   \label{r-03-01},
   \end{equation}
   \begin{equation}
   \frac{{\cal B}r(\overline{B}_{u}^{{\ast}-}{\to}D_{u}^{0}{\rho}^{-})}
        {{\cal B}r(\overline{B}_{u}^{{\ast}-}{\to}D_{u}^{0}K^{{\ast}-})}
   \ {\approx}\ \frac{f_{\rho}^{2}}{{\lambda}^{2}\,f_{K^{\ast}}^{2}}
   \label{r-03-02},
   \end{equation}
   \begin{equation}
   \frac{{\cal B}r(\overline{B}_{s}^{{\ast}0}{\to}D_{u}^{0}{\phi})}
        {{\cal B}r(\overline{B}_{s}^{{\ast}0}{\to}D_{u}^{0}K^{{\ast}0})}
   \ {\approx}\ \frac{{\lambda}^{2}\,f_{\phi}^{2}}{f_{K^{\ast}}^{2}}
   \label{r-03-03}.
   \end{equation}

  (4)
  The branching ratios for the $\overline{B}_{q}^{\ast}$ ${\to}$
  $DM$ decays are smaller by at least five orders of magnitude
  than the branching ratios for the $\overline{B}_{q}$ ${\to}$
  $DM$ decays \cite{prd78.014018}.
  This fact implies that the possible background from the
  $\overline{B}_{q}^{\ast}$ ${\to}$ $DM$ decays could be safely
  neglected when the $\overline{B}_{q}$ ${\to}$ $DM$ decays
  were analyzed, but not vice versa, i.e., one of main pollution
  backgrounds for the $\overline{B}_{q}^{\ast}$ ${\to}$ $DM$
  decays would come from the $\overline{B}_{q}$ ${\to}$ $DM$
  decays, even if the invariant mass of the $DM$ meson pair
  could be used to distinguish the $\overline{B}_{q}^{\ast}$
  meson from the $\overline{B}_{q}$ meson experimentally.

   \begin{table}[ht]
   \caption{The channel fractions at the ${\Upsilon}(5S)$ resonance \cite{epjc74.3026}.}
   \label{tab:bb-fr}
   \begin{ruledtabular}
   \begin{tabular}{ccc}
     channels & \%/$b\bar{b}$ event & \%/$B_{s}$ event \\ \hline
     All $B_{s}$ events
   & $19.5^{+3.0}_{-2.3}$ & \\
     $B_{s}^{{\ast}0}\overline{B}_{s}^{{\ast}0}$
   &  & $90.1^{+3.8}_{-4.0}{\pm}0.2$ \\
     $B_{s}^{{\ast}0}\overline{B}_{s}^{0}+{\rm c.c.}$
   &  & $7.3^{+3.3}_{-3.0}{\pm}0.1$ \\ \hline
     $B^{\ast}\overline{B}^{\ast}$
   & $37.5^{+2.1}_{-1.9}{\pm}3.0$ & \\
     $B^{\ast}\overline{B}+{\rm c.c.}$
   & $13.7{\pm}1.3{\pm}1.1$ & \\
     $B^{\ast}\overline{B}{\pi}+{\rm c.c.}$
   & $7.3^{+2.3}_{-2.1}{\pm}0.8$ & \\
     $B^{\ast}\overline{B}^{\ast}{\pi}$
   & $1.0^{+1.4}_{-1.3}{\pm}0.4$ & \\ \hline
   \end{tabular}
   \end{ruledtabular}
   \end{table}

  (5)
  The event numbers of the $B_{q}^{\ast}$ meson in a data sample
  can be calculated by the following formula,
   \begin{equation}
   N(B_{q}^{\ast})\ =\ {\cal L}_{\rm int}\,
  {\times}\, {\sigma}_{b\bar{b}}\,{\times}\,f_{B_{q}}\, {\times}\,
   \frac{ f_{B_{q}^{\ast}} } { f_{B_{q}} }
   \label{r-05-01}.
   \end{equation}
   \begin{equation}
  f_{B_{q}^{\ast}}\ =\ 2\,{\times}\,f_{B_{q}^{\ast}\overline{B}_{q}^{\ast}}
   +2\,{\times}\,f_{B_{q}^{\ast}\overline{B}_{q}^{\ast}{\pi}}
   +f_{B_{q}^{\ast}\overline{B}_{q}+{\rm c.c.}}
   +f_{B_{q}^{\ast}\overline{B}_{q}{\pi}+{\rm c.c}}
   +{\cdots}
   \label{r-02-02},
   \end{equation}
  where ${\cal L}_{\rm int}$ is the integrated luminosity,
  ${\sigma}_{b\bar{b}}$ denotes the $b\bar{b}$ pair production cross section,
  $f_{B_{q}}$, $f_{B_{q}^{\ast}\overline{B}_{q}^{\ast}}$, ${\cdots}$ refer
  to the production fraction of all the $B_{q}$ meson, the
  $B_{q}^{\ast}\overline{B}_{q}^{\ast}$ meson pair, ${\cdots}$.
  The production fractions of specific modes at the center-of-mass
  of the ${\Upsilon}(5S)$ resonance \cite{epjc74.3026} are listed
  in Table \ref{tab:bb-fr}.
  With a large production cross section of the process $e^{+}e^{-}$ ${\to}$
  $b\bar{b}$ at the ${\Upsilon}(5S)$ peak ${\sigma}_{b\bar{b}}$ $=$
  $(0.340{\pm}0.016)\,{\rm nb}$ \cite{epjc74.3026}, it is expected that
  some $3.3{\times}10^{9}$ $B_{u,d}^{\ast}$ and $1.2{\times}10^{9}$
  $B_{s}^{\ast}$ mesons could be available per $10\,{\rm ab}^{-1}$
  ${\Upsilon}(5S)$ dataset.
  The branching ratios of the color-allowed ``T-I'' class
  $\overline{B}_{q}^{\ast}$ ${\to}$ $DM$ decays can reach up
  to ${\cal O}(10^{-9}$) or more, which are essentially coincident
  with those obtained by the QCDF approach \cite{epjc76.523}.
  Hence, a few events of the $\overline{B}_{q}^{\ast}$ ${\to}$
  $D_{q}{\pi}^{-}$ and $\overline{B}_{u,d}^{\ast}$ ${\to}$
  $D_{u,d}{\rho}^{-}$ decays, and dozens of the
  $\overline{B}_{s}^{{\ast}0}$ ${\to}$ $D_{s}^{+}{\rho}^{-}$ decay,
  might be available at the forthcoming SuperKEKB.
  At high energy hadron colliders, for example, given with the
  cross section at the LHCb ${\sigma}_{b\bar{b}}$ ${\approx}$
  $100\,{\rm {\mu}b}$ \cite{pdg,crp16.435,prl118.052002},
  with a similar ratio $f_{B_{u}}$ $=$ $f_{B_{d}}$ $=$ $0.344{\pm}0.021$
  and $f_{B_{s}}$ $=$ $0.115{\pm}0.013$ at Tevatron \cite{pdg,1002.5012}
  and a similar ratio $f_{B_{q}^{\ast}}/f_{B_{q}}$ at the
  ${\Upsilon}(5S)$ meson \cite{epjc74.3026},
  some $9.8{\times}10^{13}$ $B_{u,d}^{\ast}$ events and $2.2{\times}10^{13}$
  $B_{s}^{\ast}$ events per ${\rm ab}^{-1}$ dataset could be available at
  the LHCb, corresponding to more than $10^{5}$ of the
  $\overline{B}_{s}^{{\ast}0}$ ${\to}$ $D_{s}^{+}{\rho}^{-}$ decay
  events and over $10^{4}$ of the $\overline{B}_{q}^{\ast}$ ${\to}$
  $D_{q}{\pi}^{-}$ and $\overline{B}_{u,d}^{\ast}$ ${\to}$
  $D_{u,d}{\rho}^{-}$ decay events,
  which should be easily measured by the future LHCb experiments.

  \section{Summary}
  \label{sec04}
  Besides the dominant electromagnetic decay mode, the ground vector
  $B_{q}^{\ast}$ meson ($q$ $=$ $u$, $d$ and $s$) can also decay via
  the weak interactions within the standard model.
  A large amount of the $B_{q}^{\ast}$ mesons are expected to be
  accumulated with the running LHC and the forthcoming SuperKEKB,
  which makes it seemingly possible to explore the $B_{q}^{\ast}$
  meson weak decays experimentally.
  The theoretical study is necessary to offer a ready reference.
  In this paper, we investigated the $\overline{B}_{q}^{\ast}$ ${\to}$
  $DP$, $DV$ decays with the phenomenological pQCD approach.
  It is found that the color-allowed $\overline{B}_{q}^{\ast}$ ${\to}$
  $D_{q}{\rho}^{-}$ decays have branching ratios ${\gtrsim}$ $10^{-9}$,
  and should be promisingly accessible at the high luminosity experiments
  in the future.

  \section*{Acknowledgments}
  The work is supported by the National Natural Science Foundation
  of China (Grant Nos. U1632109, 11547014 and 11475055).

  \begin{appendix}
  \section{The amplitude for the $\overline{B}_{q}^{\ast}$ ${\to}$ $DP$ decays}
  \label{amp-dp}
   \begin{equation}
  {\cal A}(B_{u}^{{\ast}-}{\to}D_{u}^{0}{\pi}^{-})\, =\,
  {\cal F}\, V_{cb}\,V_{ud}^{\ast}\, \Big\{
   \sum\limits_{i}\,{\cal M}^{T}_{i,P}
  +\sum\limits_{j}\,{\cal M}^{C}_{j,P} \Big\}
   \label{amp:bu-d0-pim},
   \end{equation}
   \begin{equation}
  {\cal A}(B_{u}^{{\ast}-}{\to}D_{u}^{0}K^{-}) \, =\,
  {\cal F}\, V_{cb}\,V_{us}^{\ast}\, \Big\{
   \sum\limits_{i}\,{\cal M}^{T}_{i,P}
  +\sum\limits_{j}\,{\cal M}^{C}_{j,P} \Big\}
   \label{amp:bu-d0-km},
   \end{equation}
   \begin{equation}
  {\cal A}(\overline{B}_{d}^{{\ast}0}{\to}D_{d}^{+}{\pi}^{-}) \, =\,
  {\cal F}\, V_{cb}\,V_{ud}^{\ast}\, \Big\{
   \sum\limits_{i}\,{\cal M}^{T}_{i,P}
  +\sum\limits_{j}\,{\cal M}^{A}_{j,P} \Big\}
   \label{amp:bd-dp-pim},
   \end{equation}
   \begin{equation}
  {\cal A}(\overline{B}_{d}^{{\ast}0}{\to}D_{d}^{+}K^{-}) \, =\,
  {\cal F}\, V_{cb}\,V_{us}^{\ast}\,
   \sum\limits_{i}\,{\cal M}^{T}_{i,P}
   \label{amp:bd-dp-km},
   \end{equation}
   \begin{equation}
   \sqrt{2}\,{\cal A}(\overline{B}_{d}^{{\ast}0}{\to}D_{u}^{0}{\pi}^{0}) \, =\,
  {\cal F}\, V_{cb}\,V_{ud}^{\ast}\, \Big\{
  -\sum\limits_{i}\,{\cal M}^{C}_{i,P}
  +\sum\limits_{j}\,{\cal M}^{A}_{j,P} \Big\}
   \label{amp:bd-du-pi0},
   \end{equation}
   \begin{equation}
   \sqrt{2}\, {\cal A}(\overline{B}_{d}^{{\ast}0}{\to}D_{u}^{0}{\eta}_{q}) \, =\,
  {\cal F}\, V_{cb}\,V_{ud}^{\ast}\, \Big\{
   \sum\limits_{i}\,{\cal M}^{C}_{i,P}
  +\sum\limits_{j}\,{\cal M}^{A}_{j,P} \Big\}
   \label{amp:bd-du-etaq},
   \end{equation}
   \begin{equation}
  {\cal A}(\overline{B}_{d}^{{\ast}0}{\to}D_{u}^{0}\overline{K}^{0}) \, =\,
  {\cal F}\, V_{cb}\,V_{us}^{\ast}\,
   \sum\limits_{i}\,{\cal M}^{C}_{i,P}
   \label{amp:bd-du-k0},
   \end{equation}
   \begin{equation}
  {\cal A}(\overline{B}_{d}^{{\ast}0}{\to}D_{s}^{+}K^{-}) \, =\,
  {\cal F}\, V_{cb}\,V_{ud}^{\ast}\,
   \sum\limits_{i}\,{\cal M}^{A}_{i,P}
   \label{amp:bd-ds-km},
   \end{equation}
   \begin{equation}
  {\cal A}(\overline{B}_{s}^{{\ast}0}{\to}D_{s}^{+}{\pi}^{-}) \, =\,
  {\cal F}\, V_{cb}\,V_{ud}^{\ast}\,
   \sum\limits_{i}\,{\cal M}^{T}_{i,P}
   \label{amp:bs-ds-pim},
   \end{equation}
   \begin{equation}
  {\cal A}(\overline{B}_{s}^{{\ast}0}{\to}D_{s}^{+}K^{-}) \, =\,
  {\cal F}\, V_{cb}\,V_{us}^{\ast}\, \Big\{
   \sum\limits_{i}\,{\cal M}^{T}_{i,P}
  +\sum\limits_{j}\,{\cal M}^{A}_{j,P} \Big\}
   \label{amp:bs-ds-km},
   \end{equation}
   \begin{equation}
  {\cal A}(\overline{B}_{s}^{{\ast}0}{\to}D_{d}^{+}{\pi}^{-}) \, =\,
  {\cal F}\, V_{cb}\,V_{us}^{\ast}\,
   \sum\limits_{i}\,{\cal M}^{A}_{i,P}
   \label{amp:bs-dd-pim},
   \end{equation}
   \begin{equation}
   \sqrt{2}\,{\cal A}(\overline{B}_{s}^{{\ast}0}{\to}D_{u}^{0}{\pi}^{0}) \, =\,
  {\cal F}\, V_{cb}\,V_{us}^{\ast}\,
   \sum\limits_{i}\,{\cal M}^{A}_{i,P}
   \label{amp:bs-du-piz},
   \end{equation}
   \begin{equation}
   \sqrt{2}\,{\cal A}(\overline{B}_{s}^{{\ast}0}{\to}D_{u}^{0}{\eta}_{q}) \, =\,
  {\cal F}\, V_{cb}\,V_{us}^{\ast}\,
   \sum\limits_{i}\,{\cal M}^{A}_{i,P}
   \label{amp:bs-du-etaq},
   \end{equation}
   \begin{equation}
  {\cal A}(\overline{B}_{s}^{{\ast}0}{\to}D_{u}^{0}{\eta}_{s}) \, =\,
  {\cal F}\, V_{cb}\,V_{us}^{\ast}\,
   \sum\limits_{i}\,{\cal M}^{C}_{i,P}
   \label{amp:bs-du-etas},
   \end{equation}
   \begin{equation}
  {\cal A}(\overline{B}_{s}^{{\ast}0}{\to}D_{u}^{0}K^{0}) \, =\,
  {\cal F}\, V_{cb}\,V_{ud}^{\ast}\,
   \sum\limits_{i}\,{\cal M}^{C}_{i,P}
   \label{amp:bs-du-kz},
   \end{equation}
   \begin{equation}
  {\cal F}\, =\, \frac{G_{F}}{\sqrt{2}}\, \frac{{\pi}\,C_{F}}{N_{c}}\,
  f_{{B}_{q}^{\ast}}\, f_{D}
   \label{eq:amp-coe},
   \end{equation}
  where ${\cal M}^{k}_{i,j}$ is the amplitude building blocks.
  The superscripts $k$ $=$ $T$, $C$, $A$ correspond to the
  color-allowed emission topologies of Fig.\ref{fig:fey-t},
  the color-suppressed emission topologies of Fig.\ref{fig:fey-c},
  the annihilation topologies of Fig.\ref{fig:fey-a}.
  The subscripts $i$ $=$ $a$, $b$, $c$, $d$ correspond to the
  diagram indices.
  The subscripts $j$ $=$ $P$, $L$, $N$, $T$ correspond to
  the different helicity amplitudes.
  The analytical expressions of the amplitude building blocks
  ${\cal M}^{k}_{i,j}$ are given in the Appendix \ref{block-t},
  \ref{block-c}, \ref{block-a}.

  \section{The amplitude for the $\overline{B}_{q}^{\ast}$ ${\to}$ $DV$ decays}
  \label{amp-dv}
   \begin{equation}
  i\,{\cal A}_{\lambda}(B_{u}^{{\ast}-}{\to}D_{u}^{0}{\rho}^{-})\, =\,
  {\cal F}\, V_{cb}\,V_{ud}^{\ast}\, \Big\{
   \sum\limits_{i}\,{\cal M}^{T}_{i,{\lambda}}
  +\sum\limits_{j}\,{\cal M}^{C}_{j,{\lambda}} \Big\}
   \label{amp:bu-d0-rhom},
   \end{equation}
   \begin{equation}
  i\,{\cal A}_{\lambda}(B_{u}^{{\ast}-}{\to}D_{u}^{0}K^{{\ast}-})\, =\,
  {\cal F}\, V_{cb}\,V_{us}^{\ast}\, \Big\{
   \sum\limits_{i}\,{\cal M}^{T}_{i,{\lambda}}
  +\sum\limits_{j}\,{\cal M}^{C}_{j,{\lambda}} \Big\}
   \label{amp:bu-d0-kvm},
   \end{equation}
   \begin{equation}
  i\,{\cal A}_{\lambda}(\overline{B}_{d}^{{\ast}0}{\to}D_{d}^{+}{\rho}^{-}) \, =\,
  {\cal F}\, V_{cb}\,V_{ud}^{\ast}\, \Big\{
   \sum\limits_{i}\,{\cal M}^{T}_{i,{\lambda}}
  +\sum\limits_{j}\,{\cal M}^{A}_{j,{\lambda}} \Big\}
   \label{amp:bd-dp-rhom},
   \end{equation}
   \begin{equation}
  i{\cal A}_{\lambda}(\overline{B}_{d}^{{\ast}0}{\to}D_{d}^{+}K^{{\ast}-}) \, =\,
  {\cal F}\, V_{cb}\,V_{us}^{\ast}\,
   \sum\limits_{i}\,{\cal M}^{T}_{i,{\lambda}}
   \label{amp:bd-dp-kvm},
   \end{equation}
   \begin{equation}
  i\,\sqrt{2}\,{\cal A}_{\lambda}(\overline{B}_{d}^{{\ast}0}{\to}D_{u}^{0}{\rho}^{0}) \, =\,
  {\cal F}\, V_{cb}\,V_{ud}^{\ast}\, \Big\{
  -\sum\limits_{i}\,{\cal M}^{C}_{i,{\lambda}}
  +\sum\limits_{j}\,{\cal M}^{A}_{j,{\lambda}} \Big\}
   \label{amp:bd-du-rhoz},
   \end{equation}
   \begin{equation}
  i\,\sqrt{2}\,{\cal A}_{\lambda}(\overline{B}_{d}^{{\ast}0}{\to}D_{u}^{0}{\omega}) \, =\,
  {\cal F}\, V_{cb}\,V_{ud}^{\ast}\, \Big\{
   \sum\limits_{i}\,{\cal M}^{C}_{i,{\lambda}}
  +\sum\limits_{j}\,{\cal M}^{A}_{j,{\lambda}} \Big\}
   \label{amp:bd-du-w},
   \end{equation}
   \begin{equation}
  i{\cal A}_{\lambda}(\overline{B}_{d}^{{\ast}0}{\to}D_{u}^{0}\overline{K}^{{\ast}0}) \, =\,
  {\cal F}\, V_{cb}\,V_{us}^{\ast}\,
   \sum\limits_{i}\,{\cal M}^{C}_{i,{\lambda}}
   \label{amp:bd-du-kvz},
   \end{equation}
   \begin{equation}
  i{\cal A}_{\lambda}(\overline{B}_{d}^{{\ast}0}{\to}D_{s}^{+}K^{{\ast}-}) \, =\,
  {\cal F}\, V_{cb}\,V_{ud}^{\ast}\,
   \sum\limits_{i}\,{\cal M}^{A}_{i,{\lambda}}
   \label{amp:bd-ds-kvm},
   \end{equation}
   \begin{equation}
  i{\cal A}_{\lambda}(\overline{B}_{s}^{{\ast}0}{\to}D_{s}^{+}{\rho}^{-}) \, =\,
  {\cal F}\, V_{cb}\,V_{ud}^{\ast}\,
   \sum\limits_{i}\,{\cal M}^{T}_{i,{\lambda}}
   \label{amp:bs-ds-rhom},
   \end{equation}
   \begin{equation}
  i{\cal A}_{\lambda}(\overline{B}_{s}^{{\ast}0}{\to}D_{s}^{+}K^{{\ast}-}) \, =\,
  {\cal F}\, V_{cb}\,V_{us}^{\ast}\, \Big\{
   \sum\limits_{i}\,{\cal M}^{T}_{i,{\lambda}}
  +\sum\limits_{j}\,{\cal M}^{A}_{j,{\lambda}} \Big\}
   \label{amp:bs-ds-kvm},
   \end{equation}
   \begin{equation}
  i{\cal A}_{\lambda}(\overline{B}_{s}^{{\ast}0}{\to}D_{d}^{+}{\rho}^{-}) \, =\,
  {\cal F}\, V_{cb}\,V_{us}^{\ast}\,
   \sum\limits_{i}\,{\cal M}^{A}_{i,{\lambda}}
   \label{amp:bs-dd-rhom},
   \end{equation}
   \begin{equation}
  i\,\sqrt{2}\,{\cal A}_{\lambda}(\overline{B}_{s}^{{\ast}0}{\to}D_{u}^{0}{\rho}^{0}) \, =\,
  {\cal F}\, V_{cb}\,V_{us}^{\ast}\,
   \sum\limits_{i}\,{\cal M}^{A}_{i,{\lambda}}
   \label{amp:bs-du-rhoz},
   \end{equation}
   \begin{equation}
  i\,\sqrt{2}\,{\cal A}_{\lambda}(\overline{B}_{s}^{{\ast}0}{\to}D_{u}^{0}{\omega}) \, =\,
  {\cal F}\, V_{cb}\,V_{us}^{\ast}\,
   \sum\limits_{i}\,{\cal M}^{A}_{i,{\lambda}}
   \label{amp:bs-du-w},
   \end{equation}
   \begin{equation}
  i{\cal A}_{\lambda}(\overline{B}_{s}^{{\ast}0}{\to}D_{u}^{0}{\phi}) \, =\,
  {\cal F}\, V_{cb}\,V_{us}^{\ast}\,
   \sum\limits_{i}\,{\cal M}^{C}_{i,{\lambda}}
   \label{amp:bs-du-phi},
   \end{equation}
   \begin{equation}
  i{\cal A}_{\lambda}(\overline{B}_{s}^{{\ast}0}{\to}D_{u}^{0}K^{{\ast}0}) \, =\,
  {\cal F}\, V_{cb}\,V_{ud}^{\ast}\,
   \sum\limits_{i}\,{\cal M}^{C}_{i,{\lambda}}
   \label{amp:bs-du-kvz},
   \end{equation}
  where the index ${\lambda}$ corresponds to three different
  helicity amplitudes, i.e., ${\lambda}$ $=$ $L$, $N$, $T$.

  \section{Amplitude building blocks for the color-allowed
   $\overline{B}_{q}^{\ast}$ ${\to}$ $D_{q}M$ decays}
  \label{block-t}
  The expressions of the amplitude building blocks ${\cal M}^{T}_{i,j}$
  for the color-allowed topologies are presented as follows, where
  the subscript $i$ corresponds to the diagram indices of Fig.\ref{fig:fey-t};
  and $j$ corresponds to the different helicity amplitudes.
   \begin{eqnarray}
  {\cal M}^{T}_{a,P} &=&
  2\,m_{1}\,p\,
  {\int}_{0}^{1}dx_{1}
  {\int}_{0}^{1}dx_{2}
  {\int}_{0}^{\infty}b_{1}db_{1}
  {\int}_{0}^{\infty}b_{2}db_{2}\,
  H^{T}_{f}({\alpha}^{T},{\beta}^{T}_{a},b_{1},b_{2})\,
  E^{T}_{f}(t^{T}_{a})
   \nonumber \\ &{\times}&
  {\alpha}_{s}(t^{T}_{a})\, a_{1}(t^{T}_{a})\,
  {\phi}_{B_{q}^{\ast}}^{v}(x_{1})\,
   \Big\{ {\phi}_{D}^{a}(x_{2})\,
   ( m_{1}^{2}\,\bar{x}_{2}+m_{3}^{2}\,x_{2} )
   + {\phi}_{D}^{p}(x_{2})\, m_{2}\,m_{b} \Big\}
   \label{amp:t-a-p},
   \end{eqnarray}
   \begin{eqnarray}
  {\cal M}^{T}_{a,L} &=&
  {\int}_{0}^{1}dx_{1}
  {\int}_{0}^{1}dx_{2}
  {\int}_{0}^{\infty}b_{1}db_{1}
  {\int}_{0}^{\infty}b_{2}db_{2}\,
  H^{T}_{f}({\alpha}^{T},{\beta}^{T}_{a},b_{1},b_{2})\,
  E^{T}_{f}(t^{T}_{a})\, {\alpha}_{s}(t^{T}_{a})
   \nonumber \\ &{\times}&
  a_{1}(t^{T}_{a})\, {\phi}_{B_{q}^{\ast}}^{v}(x_{1})\,
   \Big\{ {\phi}_{D}^{a}(x_{2})\,
   ( m_{1}^{2}\,s\,\bar{x}_{2}+m_{3}^{2}\,t\,x_{2} )
  + {\phi}_{D}^{p}(x_{2})\, m_{2}\,m_{b}\,u \Big\}
   \label{amp:t-a-l},
   \end{eqnarray}
   \begin{eqnarray}
  {\cal M}^{T}_{a,N} &=& m_{1}\,m_{3}\,
  {\int}_{0}^{1}dx_{1}
  {\int}_{0}^{1}dx_{2}
  {\int}_{0}^{\infty}b_{1}db_{1}
  {\int}_{0}^{\infty}b_{2}db_{2}\,
  H^{T}_{f}({\alpha}^{T},{\beta}^{T}_{a},b_{1},b_{2})\,
  E^{T}_{f}(t^{T}_{a})
   \nonumber \\ &{\times}&
  {\alpha}_{s}(t^{T}_{a})\, a_{1}(t^{T}_{a})\,
  {\phi}_{B_{q}^{\ast}}^{V}(x_{1})\,
   \Big\{ {\phi}_{D}^{a}(x_{2})\, (2\,m_{2}^{2}\,x_{2} -t)
  -2\, m_{2}\,m_{b}\, {\phi}_{D}^{p}(x_{2}) \Big\}
   \label{amp:t-a-n},
   \end{eqnarray}
   \begin{eqnarray}
  {\cal M}^{T}_{a,T} &=& 2\,m_{1}\,m_{3}\,
  {\int}_{0}^{1}dx_{1}
  {\int}_{0}^{1}dx_{2}
  {\int}_{0}^{\infty}b_{1}db_{1}
  {\int}_{0}^{\infty}b_{2}db_{2}\,
  H^{T}_{f}({\alpha}^{T},{\beta}^{T}_{a},b_{1},b_{2})\,
  E^{T}_{f}(t^{T}_{a})
   \nonumber \\ &{\times}&
  {\alpha}_{s}(t^{T}_{a})\, a_{1}(t^{T}_{a})\,
  {\phi}_{B_{q}^{\ast}}^{V}(x_{1})\, {\phi}_{D}^{a}(x_{2})
   \label{amp:t-a-t},
   \end{eqnarray}
   \begin{eqnarray}
  {\cal M}^{T}_{b,P} &=&
  2\,m_{1}\,p\,
  {\int}_{0}^{1}dx_{1}
  {\int}_{0}^{1}dx_{2}
  {\int}_{0}^{\infty}b_{1}db_{1}
  {\int}_{0}^{\infty}b_{2}db_{2}\,
  H^{T}_{f}({\alpha}^{T},{\beta}^{T}_{b},b_{2},b_{1})\,
  E^{T}_{f}(t^{T}_{b})
   \nonumber \\ &{\times}&
  {\alpha}_{s}(t^{T}_{b})\, \Big\{
  {\phi}_{B_{q}^{\ast}}^{v}(x_{1})\, \Big[
  2\,m_{2}\,m_{c}\,{\phi}_{D}^{p}(x_{2})
 -{\phi}_{D}^{a}(x_{2})\, (m_{2}^{2}\,
  \bar{x}_{1}+m_{3}^{2}\,x_{1}) \Big]
   \nonumber \\ &+&
  {\phi}_{B_{q}^{\ast}}^{t}(x_{1})\, \Big[
   2\,m_{1}\,m_{2}\,{\phi}_{D}^{p}(x_{2})\, \bar{x}_{1}
  -m_{1}\, m_{c}\, {\phi}_{D}^{a}(x_{2}) \Big] \Big\}\, a_{1}(t^{T}_{b})
   \label{amp:t-b-p},
   \end{eqnarray}
   \begin{eqnarray}
  {\cal M}^{T}_{b,L} &=&
  {\int}_{0}^{1}dx_{1}
  {\int}_{0}^{1}dx_{2}
  {\int}_{0}^{\infty}b_{1}db_{1}
  {\int}_{0}^{\infty}b_{2}db_{2}\,
  H^{T}_{f}({\alpha}^{T},{\beta}^{T}_{b},b_{2},b_{1})\,
  E^{T}_{f}(t^{T}_{b})\, {\alpha}_{s}(t^{T}_{b})
   \nonumber \\ &{\times}&
   a_{1}(t^{T}_{b})\, \Big\{
  {\phi}_{B_{q}^{\ast}}^{t}(x_{1})\, \Big[
  2\,m_{1}\,m_{2}\, {\phi}_{D}^{p}(x_{2})\,(s-u\,x_{1})
  -m_{1}\,m_{c}\,s\,{\phi}_{D}^{a}(x_{2}) \Big]
   \nonumber \\ &+&
  {\phi}_{B_{q}^{\ast}}^{v}(x_{1})\, \Big[
  {\phi}_{D}^{a}(x_{2})\,( m_{3}^{2}\,t\,\,x_{1}
  -m_{2}^{2}\,u\,\bar{x}_{1} )
  +2\,m_{2}\,m_{c}\,u\,{\phi}_{D}^{p}(x_{2}) \Big] \Big\}
   \label{amp:t-b-l},
   \end{eqnarray}
   \begin{eqnarray}
  {\cal M}^{T}_{b,N} &=& m_{3}\,
  {\int}_{0}^{1}dx_{1}
  {\int}_{0}^{1}dx_{2}
  {\int}_{0}^{\infty}b_{1}db_{1}
  {\int}_{0}^{\infty}b_{2}db_{2}\,
  H^{T}_{f}({\alpha}^{T},{\beta}^{T}_{b},b_{2},b_{1})\,
  E^{T}_{f}(t^{T}_{b})
   \nonumber \\ &{\times}&
 {\alpha}_{s}(t^{T}_{b})\, \Big\{
  {\phi}_{B_{q}^{\ast}}^{V}(x_{1})\, m_{1}\,\Big[
  {\phi}_{D}^{a}(x_{2})\,(2\,m_{2}^{2}-t\,x_{1})
  -4\,m_{2}\,m_{c}\,{\phi}_{D}^{p}(x_{2}) \Big]
   \nonumber \\ &+&
  {\phi}_{B_{q}^{\ast}}^{T}(x_{1})\, \Big[
  {\phi}_{D}^{a}(x_{2})\,t\,m_{c}
 +{\phi}_{D}^{p}(x_{2})\,2\,m_{2}\,(2\,m_{1}^{2}\,x_{1}-t)
   \Big] \Big\}\, a_{1}(t^{T}_{b})
   \label{amp:t-b-n},
   \end{eqnarray}
   \begin{eqnarray}
  {\cal M}^{T}_{b,T} &=& 2\,m_{3}\,
  {\int}_{0}^{1}dx_{1}
  {\int}_{0}^{1}dx_{2}
  {\int}_{0}^{\infty}b_{1}db_{1}
  {\int}_{0}^{\infty}b_{2}db_{2}\,
  H^{T}_{f}({\alpha}^{T},{\beta}^{T}_{b},b_{2},b_{1})\,
  E^{T}_{f}(t^{T}_{b})\, {\alpha}_{s}(t^{T}_{b})
   \nonumber \\ &{\times}&
  a_{1}(t^{T}_{b})\, \Big\{
  {\phi}_{B_{q}^{\ast}}^{T}(x_{1})\, \Big[
  {\phi}_{D}^{p}(x_{2})\,2\,m_{2}
 -{\phi}_{D}^{a}(x_{2})\,m_{c} \Big]
 -{\phi}_{B_{q}^{\ast}}^{V}(x_{1})\,
  {\phi}_{D}^{a}(x_{2})\,m_{1}\,x_{1} \Big\}
   \label{amp:t-b-t},
   \end{eqnarray}
   \begin{eqnarray}
  {\cal M}^{T}_{c,P} &=&
   \frac{2\,m_{1}\,p}{N_{c}}\,
  {\int}_{0}^{1}dx_{1}
  {\int}_{0}^{1}dx_{2}
  {\int}_{0}^{1}dx_{3}
  {\int}_{0}^{\infty}db_{1}
  {\int}_{0}^{\infty}b_{2}db_{2}
  {\int}_{0}^{\infty}b_{3}db_{3}\,
  {\delta}(b_{1}-b_{2})
   \nonumber \\ &{\times}&
  {\phi}_{P}^{a}(x_{3})\,  {\alpha}_{s}(t^{T}_{c})\,
  C_{2}(t^{T}_{c})\,
  \Big\{ {\phi}_{B_{q}^{\ast}}^{v}(x_{1})\, {\phi}_{D}^{a}(x_{2})\,
  (2\,m_{2}^{2}\,x_{2}+s\,\bar{x}_{3}-t\,x_{1})
   \nonumber \\ &+&
  {\phi}_{B_{q}^{\ast}}^{t}(x_{1})\, {\phi}_{D}^{p}(x_{2})\,
   m_{1}\,m_{2}\,(x_{1}-x_{2}) \Big\}\,
   H^{T}_{n}({\alpha}^{T},{\beta}^{T}_{c},b_{3},b_{2})\,
   E_{n}(t^{T}_{c})
   \label{amp:t-c-p},
   \end{eqnarray}
   \begin{eqnarray}
  {\cal M}^{T}_{c,L} &=&
  \frac{1}{N_{c}}\,
  {\int}_{0}^{1}dx_{1}
  {\int}_{0}^{1}dx_{2}
  {\int}_{0}^{1}dx_{3}
  {\int}_{0}^{\infty}db_{1}
  {\int}_{0}^{\infty}b_{2}db_{2}
  {\int}_{0}^{\infty}b_{3}db_{3}\,
  H^{T}_{n}({\alpha}^{T},{\beta}^{T}_{c},b_{3},b_{2})
   \nonumber \\ &{\times}&
   {\delta}(b_{1}-b_{2})\,
   E_{n}(t^{T}_{c})\, {\phi}_{V}^{v}(x_{3})\,
   \Big\{ {\phi}_{B_{q}^{\ast}}^{v}(x_{1})\,
  {\phi}_{D}^{a}(x_{2})\, u\,
   (2\,m_{2}^{2}\,x_{2}+s\,\bar{x}_{3}-t\,x_{1})
   \nonumber \\ &+&
   {\phi}_{B_{q}^{\ast}}^{t}(x_{1})\,
   {\phi}_{D}^{p}(x_{2})\, m_{1}\,m_{2}\,
   (u\,x_{1}-s\,x_{2}-2\,m_{3}^{2}\,\bar{x}_{3})
   \Big\}\,
  {\alpha}_{s}(t^{T}_{c})\,
   C_{2}(t^{T}_{c})
   \label{amp:t-c-l},
   \end{eqnarray}
   \begin{eqnarray}
  {\cal M}^{T}_{c,N} &=&
   \frac{m_{3}}{N_{c}}\,
  {\int}_{0}^{1}dx_{1}
  {\int}_{0}^{1}dx_{2}
  {\int}_{0}^{1}dx_{3}
  {\int}_{0}^{\infty}db_{1}
  {\int}_{0}^{\infty}b_{2}db_{2}
  {\int}_{0}^{\infty}b_{3}db_{3}
   \nonumber \\ &{\times}&
  H^{T}_{n}({\alpha}^{T},{\beta}^{T}_{c},b_{3},b_{2})\,
  E_{n}(t^{T}_{c})\, {\alpha}_{s}(t^{T}_{c})\,
  C_{2}(t^{T}_{c})\, {\delta}(b_{1}-b_{2})
   \nonumber \\ &{\times}&
   \Big\{ {\phi}_{B_{q}^{\ast}}^{V}(x_{1})\, {\phi}_{D}^{a}(x_{2})\,
  {\phi}_{V}^{V}(x_{3})\, 2\,m_{1}\,
  (t\,x_{1}-2\,m_{2}^{2}\,x_{2}-s\,\bar{x}_{3})
   \nonumber \\ & & +
  {\phi}_{B_{q}^{\ast}}^{T}(x_{1})\, {\phi}_{D}^{p}(x_{2})\,
  {\phi}_{V}^{V}(x_{3})\, m_{2}\,
  (t\,x_{2}+u\,\bar{x}_{3}-2\,m_{1}^{2}\,x_{1})
   \nonumber \\ & & +
  {\phi}_{B_{q}^{\ast}}^{T}(x_{1})\, {\phi}_{D}^{p}(x_{2})\,
  {\phi}_{V}^{A}(x_{3})\, 2\,m_{1}\,m_{2}\,p\,
  (x_{2}-\bar{x}_{3}) \Big\}
   \label{amp:a-03},
   \end{eqnarray}
   \begin{eqnarray}
  {\cal M}^{T}_{c,T} &=&
   \frac{m_{3}}{N_{c}\,p}\,
  {\int}_{0}^{1}dx_{1}
  {\int}_{0}^{1}dx_{2}
  {\int}_{0}^{1}dx_{3}
  {\int}_{0}^{\infty}db_{1}
  {\int}_{0}^{\infty}b_{2}db_{2}
  {\int}_{0}^{\infty}b_{3}db_{3}
   \nonumber \\ &{\times}&
  H^{T}_{n}({\alpha}^{T},{\beta}^{T}_{c},b_{3},b_{2})\,
  E_{n}(t^{T}_{c})\, {\alpha}_{s}(t^{T}_{c})\,
  C_{2}(t^{T}_{c})\, {\delta}(b_{1}-b_{2})
   \nonumber \\ &{\times}&
   \Big\{ {\phi}_{B_{q}^{\ast}}^{V}(x_{1})\, {\phi}_{D}^{a}(x_{2})\,
  {\phi}_{V}^{A}(x_{3})\, 2\, (2\,m_{2}^{2}\,x_{2}+s\,\bar{x}_{3}-t\,x_{1})
   \nonumber \\ & & +
  {\phi}_{B_{q}^{\ast}}^{T}(x_{1})\, {\phi}_{D}^{p}(x_{2})\,
  {\phi}_{V}^{A}(x_{3})\, r_{2}\,
  (2\,m_{1}^{2}\,x_{1}-t\,x_{2}-u\,\bar{x}_{3})
   \nonumber \\ & & +
  {\phi}_{B_{q}^{\ast}}^{T}(x_{1})\, {\phi}_{D}^{p}(x_{2})\,
  {\phi}_{V}^{V}(x_{3})\, 2\,m_{2}\,p\, (\bar{x}_{3}-x_{2}) \Big\}
   \label{amp:t-c-t},
   \end{eqnarray}
   \begin{eqnarray}
  {\cal M}^{T}_{d,P} &=&
   \frac{2\,m_{1}\,p}{N_{c}}\,
  {\int}_{0}^{1}dx_{1}
  {\int}_{0}^{1}dx_{2}
  {\int}_{0}^{1}dx_{3}
  {\int}_{0}^{\infty}db_{1}
  {\int}_{0}^{\infty}b_{2}db_{2}
  {\int}_{0}^{\infty}b_{3}db_{3}\,
  {\phi}_{P}^{a}(x_{3})
   \nonumber \\ &{\times}&
  {\delta}(b_{1}-b_{2})\, {\alpha}_{s}(t^{T}_{d})\,
  C_{2}(t^{T}_{d})\, E_{n}(t^{T}_{d})\, \Big\{
  {\phi}_{B_{q}^{\ast}}^{v}(x_{1})\,
 {\phi}_{D}^{a}(x_{2})\, s\,(x_{2}-x_{3})
  \nonumber \\ &+&
 {\phi}_{B_{q}^{\ast}}^{t}(x_{1})\, {\phi}_{D}^{p}(x_{2})\,
   m_{1}\,m_{2}\,(x_{1}-x_{2}) \Big\}\,
  H^{T}_{n}({\alpha}^{T},{\beta}^{T}_{d},b_{3},b_{2})
   \label{amp:t-d-p},
   \end{eqnarray}
   \begin{eqnarray}
  {\cal M}^{T}_{d,L} &=&
   \frac{1}{N_{c}}\,
  {\int}_{0}^{1}dx_{1}
  {\int}_{0}^{1}dx_{2}
  {\int}_{0}^{1}dx_{3}
  {\int}_{0}^{\infty}db_{1}
  {\int}_{0}^{\infty}b_{2}db_{2}
  {\int}_{0}^{\infty}b_{3}db_{3}\,
  {\delta}(b_{1}-b_{2})
   \nonumber \\ &{\times}&
  E_{n}(t^{T}_{d})\, {\alpha}_{s}(t^{T}_{d})\,
  C_{2}(t^{T}_{d})\, {\phi}_{V}^{v}(x_{3})\,
   \Big\{ {\phi}_{B_{q}^{\ast}}^{v}(x_{1})\, {\phi}_{D}^{a}(x_{2})\,
   4\,m_{1}^{2}\,p^{2}\,(x_{2}-x_{3})
   \nonumber \\ &+&
  {\phi}_{B_{q}^{\ast}}^{t}(x_{1})\, {\phi}_{D}^{p}(x_{2})\,
   m_{1}\,m_{2}\,( u\,x_{1}-s\,x_{2}-2\,m_{3}^{2}\,x_{3}) \Big\}\,
  H^{T}_{n}({\alpha}^{T},{\beta}^{T}_{d},b_{3},b_{2})
   \label{amp:t-d-l},
   \end{eqnarray}
   \begin{eqnarray}
  {\cal M}^{T}_{d,N} &=&
   \frac{m_{2}\,m_{3}}{N_{c}}\,
  {\int}_{0}^{1}dx_{1}
  {\int}_{0}^{1}dx_{2}
  {\int}_{0}^{1}dx_{3}
  {\int}_{0}^{\infty}db_{1}
  {\int}_{0}^{\infty}b_{2}db_{2}
  {\int}_{0}^{\infty}b_{3}db_{3}\,
  {\delta}(b_{1}-b_{2})
   \nonumber \\ &{\times}&
  {\phi}_{B_{q}^{\ast}}^{T}(x_{1})\,
  {\phi}_{D}^{p}(x_{2})\, {\alpha}_{s}(t^{T}_{d})\,
  C_{2}(t^{T}_{d})\, \Big\{ {\phi}_{V}^{V}(x_{3})\,
  ( t\,x_{2}+u\,x_{3}-2\,m_{1}^{2}\,x_{1})
   \nonumber \\ &+&
  {\phi}_{V}^{A}(x_{3})\,2\,m_{1}\,p\,(x_{2}-x_{3})
   \Big\}\, H^{T}_{n}({\alpha}^{T},{\beta}^{T}_{d},b_{3},b_{2})\,
   E_{n}(t^{T}_{d})
   \label{amp:t-d-n},
   \end{eqnarray}
   \begin{eqnarray}
  {\cal M}^{T}_{d,T} &=&
   \frac{m_{2}\,m_{3}}{N_{c}\,m_{1}\,p}\,
  {\int}_{0}^{1}dx_{1}
  {\int}_{0}^{1}dx_{2}
  {\int}_{0}^{1}dx_{3}
  {\int}_{0}^{\infty}db_{1}
  {\int}_{0}^{\infty}b_{2}db_{2}
  {\int}_{0}^{\infty}b_{3}db_{3}\,
  {\delta}(b_{1}-b_{2})
   \nonumber \\ &{\times}&
  {\phi}_{B_{q}^{\ast}}^{T}(x_{1})\,
  {\phi}_{D}^{p}(x_{2})\, {\alpha}_{s}(t^{T}_{d})\,
  C_{2}(t^{T}_{d})\, \Big\{ {\phi}_{V}^{A}(x_{3})\,
  ( 2\,m_{1}^{2}\,x_{1}-t\,x_{2}-u\,x_{3})
   \nonumber \\ &+&
  {\phi}_{V}^{V}(x_{3})\,2\,m_{1}\,p\,(x_{3}-x_{2})
   \Big\}\, H^{T}_{n}({\alpha}^{T},{\beta}^{T}_{d},b_{3},b_{2})\,
   E_{n}(t^{T}_{d})
   \label{amp:t-d-t},
   \end{eqnarray}
  where $N_{c}$ $=$ $3$ is the color number.
  ${\alpha}_{s}$ is the strong coupling constant.
  $C_{1,2}$ are the Wilson coefficients.
  The parameter $a_{i}$ is defined as
   \begin{eqnarray}
   a_{1} &=& C_{1}+\frac{1}{N_{c}}\,C_{2}
   \label{coe-a1}, \\
   a_{2} &=& C_{2}+\frac{1}{N_{c}}\,C_{1}
   \label{coe-a2}.
   \end{eqnarray}

  The functions $H_{f,n}^{T}$ and the Sudakov factors $E_{f,n}^{T}$
  are defined as follows, where the subscripts $f$ and $n$ correspond
  to the factorizable and nonfactorizable topologies.
   \begin{equation}
   H_{f}^{T}({\alpha},{\beta},b_{i},b_{j})\, =\,
   K_{0}(b_{i}\sqrt{-{\alpha}})\, \Big\{
   {\theta}(b_{i}-b_{j}) K_{0}(b_{i}\sqrt{-{\beta}})\,
   I_{0}(b_{j}\sqrt{-{\beta}})
   + (b_{i} {\leftrightarrow} b_{j}) \Big\}
   \label{amp:hft},
   \end{equation}
   \begin{eqnarray}
   H_{n}^{T}({\alpha},{\beta},b_{i},b_{j}) &=&
   \Big\{ {\theta}(b_{i}-b_{j})\, K_{0}(b_{i}\sqrt{-{\alpha}})\,
   I_{0}(b_{j}\sqrt{-{\alpha}}) + (b_{i} {\leftrightarrow} b_{j})
   \Big\}
   \nonumber \\ & &  \!\!\!\!\!\!\!\!\!\!\!\!\!\!\!\!\!\!\!\!
   \!\!\!\!\!\!\!\!\!\!\!\!\!\!\!\! {\times}\,
   \Big\{ {\theta}(-{\beta})\, K_{0}(b_{i}\sqrt{-{\beta}})
  +\frac{{\pi}}{2}\,
  {\theta}(+{\beta})\, \Big[ i\,J_{0}(b_{i}\sqrt{{\beta}})
   - Y_{0}(b_{i}\sqrt{{\beta}}) \Big] \Big\}
   \label{amp:hnt},
   \end{eqnarray}
   \begin{equation}
   E_{f}^{T}(t)\ =\ {\exp}\{ -S_{B_{q}^{\ast}}(t)-S_{D}(t) \}
   \label{sudakov-ft},
   \end{equation}
   \begin{equation}
   E_{n}(t)\ =\ {\exp}\{ -S_{B_{q}^{\ast}}(t)-S_{D}(t)-S_{M}(t) \}
   \label{sudakov-nt},
   \end{equation}
   \begin{equation}
   S_{B_{q}^{\ast}}(t)\, =\, s(x_{1},b_{1},p_{1}^{+})
  +2{\int}_{1/b_{1}}^{t}\frac{d{\mu}}{\mu}{\gamma}_{q}
   \label{sudakov-bq},
   \end{equation}
   \begin{equation}
  S_{D}(t)\, =\, s(x_{2},b_{2},p_{2}^{+}) + s(\bar{x}_{2},b_{2},p_{2}^{+})
  +2{\int}_{1/b_{2}}^{t}\frac{d{\mu}}{\mu}{\gamma}_{q}
   \label{sudakov-cq},
   \end{equation}
   \begin{equation}
  S_{M}(t)\, =\, s(x_{3},b_{3},p_{3}^{+}) + s(\bar{x}_{3},b_{3},p_{3}^{+})
  +2{\int}_{1/b_{3}}^{t}\frac{d{\mu}}{\mu}{\gamma}_{q}
   \label{sudakov-m},
   \end{equation}
  where $I_{0}$, $J_{0}$, $K_{0}$ and $Y_{0}$ are the Bessel
  functions; ${\gamma}_{q}$ $=$ $-{\alpha}_{s}/{\pi}$ is the
  quark anomalous dimension; the expression of $s(x,b,Q)$
  can be found in the appendix of Ref.\cite{prd52.3958};
  ${\alpha}^{T}$ and ${\beta}_{i}^{T}$ are the virtualities
  of the gluon and quark propagators;
  the subscripts of the quark virtuality ${\beta}_{i}^{T}$
  and the typical scale $t_{i}^{T}$ correspond to the diagram
  indices of Fig.\ref{fig:fey-t}.
   \begin{eqnarray}
  {\alpha}^{T} &=& x_{1}^{2}\,m_{1}^{2}+x_{2}^{2}\,m_{2}^{2}-x_{1}\,x_{2}\,t
   \label{gluon-t}, \\
  {\beta}_{a}^{T} &=& x_{2}^{2}\,m_{2}^{2}-x_{2}\,t+m_{1}^{2}-m_{b}^{2}
   \label{beta-ta}, \\
  {\beta}_{b}^{T} &=& x_{1}^{2}\,m_{1}^{2}-x_{1}\,t+m_{2}^{2}-m_{c}^{2}
   \label{beta-tb}, \\
  {\beta}_{c}^{T} &=& {\alpha}^{T}+\bar{x}_{3}^{2}\,m_{3}^{2}
                     -x_{1}\,\bar{x}_{3}\,u+x_{2}\,\bar{x}_{3}\,s
   \label{beta-tc}, \\
  {\beta}_{d}^{T} &=& {\alpha}^{T}+x_{3}^{2}\,m_{3}^{2}
                     -x_{1}\,x_{3}\,u+x_{2}\,x_{3}\,s
   \label{beta-td}, \\
   t_{a(b)}^{T} &=&
  {\max}(\sqrt{-{\alpha}^{T}},\sqrt{{\vert}{\beta}_{a(b)}^{T}{\vert}},1/b_{1},1/b_{2})
   \label{t-tab}, \\
   t_{c(d)}^{T} &=&
  {\max}(\sqrt{-{\alpha}^{T}},\sqrt{{\vert}{\beta}_{c(d)}^{T}{\vert}},1/b_{2},1/b_{3})
   \label{t-tcd}.
   \end{eqnarray}

  \section{Amplitude building blocks for the color-suppressed
    $\overline{B}_{q}^{\ast}$ ${\to}$ $DM_{q}$ decays}
  \label{block-c}
  The expressions of the amplitude building blocks ${\cal M}^{C}_{i,j}$
  for the color-suppressed topologies are displayed as follows, where
  the subscript $i$ corresponds to the diagram indices of Fig.\ref{fig:fey-c};
  and $j$ corresponds to the different helicity amplitudes.
   \begin{eqnarray}
  {\cal M}^{C}_{a,P} &=&
  -{\int}_{0}^{1}dx_{1}
  {\int}_{0}^{1}dx_{3}
  {\int}_{0}^{\infty}b_{1}db_{1}
  {\int}_{0}^{\infty}b_{3}db_{3}\,
  {\phi}_{B_{q}^{\ast}}^{v}(x_{1})\,
  {\alpha}_{s}(t^{C}_{a})\, a_{2}(t^{C}_{a})
   \nonumber \\ &{\times}&
   H^{C}_{f}({\alpha}^{C},{\beta}^{C}_{a},b_{1},b_{3})\,
   \Big\{ 2\,m_{1}\,p\, {\phi}_{P}^{a}(x_{3})\,
   (m_{1}^{2}\,\bar{x}_{3}+m_{2}^{2}\,x_{3})
   \nonumber \\ &+&
   2\,m_{1}\,p\, {\mu}_{P}\,m_{b}\, {\phi}_{P}^{p}(x_{3})
   + {\mu}_{P}\,m_{b}\, t\, {\phi}_{P}^{t}(x_{3}) \Big\}\,
   E^{C}_{f}(t^{C}_{a})
   \label{amp:c-a-p},
   \end{eqnarray}
   \begin{eqnarray}
  {\cal M}^{C}_{a,L} &=&
 -{\int}_{0}^{1}dx_{1}
  {\int}_{0}^{1}dx_{3}
  {\int}_{0}^{\infty}b_{1}db_{1}
  {\int}_{0}^{\infty}b_{3}db_{3}\,
  H^{C}_{f}({\alpha}^{C},{\beta}^{C}_{a},b_{1},b_{3})
   \nonumber \\ &{\times}&
  {\alpha}_{s}(t^{C}_{a})\, a_{2}(t^{C}_{a})\,
  {\phi}_{B_{q}^{\ast}}^{v}(x_{1})\,
   \Big\{ {\phi}_{V}^{v}(x_{3})\,
  (m_{1}^{2}\,s\,\bar{x}_{3}+m_{2}^{2}\,u\,x_{3})
   \nonumber \\ &+&
   m_{3}\,m_{b}\,t\, {\phi}_{V}^{t}(x_{3})
   + 2\,m_{1}\,p\, m_{3}\,m_{b}\, {\phi}_{V}^{s}(x_{3})
   \Big\}\,  E^{C}_{f}(t^{C}_{a})
   \label{amp:c-a-l},
   \end{eqnarray}
   \begin{eqnarray}
  {\cal M}^{C}_{a,N} &=&
  {\int}_{0}^{1}dx_{1}
  {\int}_{0}^{1}dx_{3}
  {\int}_{0}^{\infty}b_{1}db_{1}
  {\int}_{0}^{\infty}b_{3}db_{3}\,
  H^{C}_{f}({\alpha}^{C},{\beta}^{C}_{a},b_{1},b_{3})
   \nonumber \\ &{\times}&
  {\alpha}_{s}(t^{C}_{a})\, a_{2}(t^{C}_{a})\,
  {\phi}_{B_{q}^{\ast}}^{V}(x_{1})\,
   \Big\{ {\phi}_{V}^{V}(x_{3})\,m_{1}\,m_{3}\,(t-s\,x_{3})
   \nonumber \\ &+&
   m_{1}\,m_{b}\,s\, {\phi}_{V}^{T}(x_{3})
   + 2\,m_{3}\,p\, m_{1}^{2}\,\bar{x}_{3}\, {\phi}_{V}^{A}(x_{3})
   \Big\}\,  E^{C}_{f}(t^{C}_{a})
   \label{amp:c-a-n},
   \end{eqnarray}
   \begin{eqnarray}
  {\cal M}^{C}_{a,T} &=&
 -{\int}_{0}^{1}dx_{1}
  {\int}_{0}^{1}dx_{3}
  {\int}_{0}^{\infty}b_{1}db_{1}
  {\int}_{0}^{\infty}b_{3}db_{3}\,
  H^{C}_{f}({\alpha}^{C},{\beta}^{C}_{a},b_{1},b_{3})
   \nonumber \\ &{\times}&
  {\alpha}_{s}(t^{C}_{a})\, a_{2}(t^{C}_{a})\,
  {\phi}_{B_{q}^{\ast}}^{V}(x_{1})\,
   \Big\{ (m_{3}/p)\, {\phi}_{V}^{A}(x_{3})\,(t-s\,x_{3})
   \nonumber \\ &+&
   {\phi}_{V}^{V}(x_{3})\,2\,m_{1}\,m_{3}\,\bar{x}_{3}
  + {\phi}_{V}^{T}(x_{3})\,2\,m_{1}\,m_{b}
   \Big\}\,  E^{C}_{f}(t^{C}_{a})
   \label{amp:c-a-t},
   \end{eqnarray}
   \begin{eqnarray}
  {\cal M}^{C}_{b,P} &=&
  2\,m_{1}\,p\,
  {\int}_{0}^{1}dx_{1}
  {\int}_{0}^{1}dx_{3}
  {\int}_{0}^{\infty}b_{1}db_{1}
  {\int}_{0}^{\infty}b_{3}db_{3}\,
  H^{C}_{f}({\alpha}^{C},{\beta}^{C}_{b},b_{3},b_{1})\,
  E^{C}_{f}(t^{C}_{b})\, {\alpha}_{s}(t^{C}_{b})
   \nonumber \\ &{\times}&
  a_{2}(t^{C}_{b})\, \Big\{ {\phi}_{B_{q}^{\ast}}^{v}(x_{1})\,
  {\phi}_{P}^{a}(x_{3})\, (m_{3}^{2}\,\bar{x}_{1}+m_{2}^{2}\,x_{1})
  -{\phi}_{B_{q}^{\ast}}^{t}(x_{1})\, {\phi}_{P}^{p}(x_{3})\,
  2\,m_{1}\,{\mu}_{P}\,\bar{x}_{1} \Big\}
   \label{amp:c-b-p},
   \end{eqnarray}
   \begin{eqnarray}
  {\cal M}^{C}_{b,L} &=&
  {\int}_{0}^{1}dx_{1}
  {\int}_{0}^{1}dx_{3}
  {\int}_{0}^{\infty}b_{1}db_{1}
  {\int}_{0}^{\infty}b_{3}db_{3}\,
  H^{C}_{f}({\alpha}^{C},{\beta}^{C}_{b},b_{3},b_{1})\,
  E^{C}_{f}(t^{C}_{b})\, {\alpha}_{s}(t^{C}_{b})\,
  a_{2}(t^{C}_{b})
   \nonumber \\ &{\times}&
   \Big\{ {\phi}_{B_{q}^{\ast}}^{v}(x_{1})\,
  {\phi}_{V}^{v}(x_{3})\, (m_{2}^{2}\,u\,x_{1}-m_{3}^{2}\,t\,\bar{x}_{1})
  -{\phi}_{B_{q}^{\ast}}^{t}(x_{1})\, {\phi}_{V}^{s}(x_{3})\,
  4\,m_{1}^{2}\,m_{3}\,p\,\bar{x}_{1} \Big\}
   \label{amp:c-b-l},
   \end{eqnarray}
   \begin{eqnarray}
  {\cal M}^{C}_{b,N} &=&
  m_{1}\,m_{3}\, {\int}_{0}^{1}dx_{1}
  {\int}_{0}^{1}dx_{3}
  {\int}_{0}^{\infty}b_{1}db_{1}
  {\int}_{0}^{\infty}b_{3}db_{3}\,
  H^{C}_{f}({\alpha}^{C},{\beta}^{C}_{b},b_{3},b_{1})\,
  E^{C}_{f}(t^{C}_{b})
   \nonumber \\ &{\times}&
  {\alpha}_{s}(t^{C}_{b})\, a_{2}(t^{C}_{b})\,
  {\phi}_{B_{q}^{\ast}}^{V}(x_{1})\,  \Big\{
  {\phi}_{V}^{V}(x_{3})\, (s-t\,x_{1})
 +{\phi}_{V}^{A}(x_{3})\, 2\,m_{1}\,p\,\bar{x}_{1} \Big\}
   \label{amp:c-b-n},
   \end{eqnarray}
   \begin{eqnarray}
  {\cal M}^{C}_{b,T} &=&
   \frac{-m_{3}}{p}\, {\int}_{0}^{1}dx_{1}
  {\int}_{0}^{1}dx_{3}
  {\int}_{0}^{\infty}b_{1}db_{1}
  {\int}_{0}^{\infty}b_{3}db_{3}\,
  H^{C}_{f}({\alpha}^{C},{\beta}^{C}_{b},b_{3},b_{1})\,
  E^{C}_{f}(t^{C}_{b})
   \nonumber \\ &{\times}&
  {\alpha}_{s}(t^{C}_{b})\, a_{2}(t^{C}_{b})\,
  {\phi}_{B_{q}^{\ast}}^{V}(x_{1})\,  \Big\{
  {\phi}_{V}^{A}(x_{3})\, (s-t\,x_{1})
 +{\phi}_{V}^{V}(x_{3})\, 2\,m_{1}\,p\,\bar{x}_{1} \Big\}
   \label{amp:c-b-t},
   \end{eqnarray}
   \begin{eqnarray}
  {\cal M}^{C}_{c,P} &=&
  {\int}_{0}^{1}dx_{1}
  {\int}_{0}^{1}dx_{2}
  {\int}_{0}^{1}dx_{3}
  {\int}_{0}^{\infty}db_{1}
  {\int}_{0}^{\infty}b_{2}db_{2}
  {\int}_{0}^{\infty}b_{3}db_{3}\,
  H^{C}_{n}({\alpha}^{C},{\beta}^{C}_{c},b_{2},b_{3})
   \nonumber \\ &{\times}&
  {\delta}(b_{1}-b_{3})\,  \Big\{
  {\phi}_{B_{q}^{\ast}}^{t}(x_{1})\, {\phi}_{D}^{a}(x_{2})\,
   m_{1}\, {\mu}_{P}\, \Big[ {\phi}_{P}^{t}(x_{3})\,
  (t\,x_{1}-2\,m_{2}^{2}\,\bar{x}_{2}-s\,x_{3})
  \nonumber \\ &+&
  {\phi}_{P}^{p}(x_{3})\, 2\,m_{1}\,p\,(x_{3}-x_{1})
  \Big] -
  {\phi}_{B_{q}^{\ast}}^{v}(x_{1})\, {\phi}_{P}^{a}(x_{3})\,
  2\,m_{1}\,p\,\Big[ {\phi}_{D}^{p}(x_{2})\, m_{2}\,m_{c}
   \nonumber \\ &+&
  {\phi}_{D}^{a}(x_{2})\, (s\,\bar{x}_{2}+2\,m_{3}^{2}\,x_{3}-u\,x_{1})
   \Big] \Big\}\, E_{n}(t^{C}_{c})\, {\alpha}_{s}(t^{C}_{c})\,
   C_{1}(t^{C}_{c})/N_{c}
   \label{amp:c-c-p},
   \end{eqnarray}
   \begin{eqnarray}
  {\cal M}^{C}_{c,L} &=&
  {\int}_{0}^{1}dx_{1}
  {\int}_{0}^{1}dx_{2}
  {\int}_{0}^{1}dx_{3}
  {\int}_{0}^{\infty}db_{1}
  {\int}_{0}^{\infty}b_{2}db_{2}
  {\int}_{0}^{\infty}b_{3}db_{3}\,
  H^{C}_{n}({\alpha}^{C},{\beta}^{C}_{c},b_{2},b_{3})
   \nonumber \\ &{\times}&
   {\delta}(b_{1}-b_{3})\,  \Big\{
  {\phi}_{B_{q}^{\ast}}^{t}(x_{1})\, {\phi}_{D}^{a}(x_{2})\,
   m_{1}\, m_{3}\, \Big[ {\phi}_{V}^{t}(x_{3})\,
  (t\,x_{1}-2\,m_{2}^{2}\,\bar{x}_{2}-s\,x_{3})
  \nonumber \\ &+&
  {\phi}_{V}^{s}(x_{3})\, 2\,m_{1}\,p\,(x_{3}-x_{1})
   \Big] + {\phi}_{B_{q}^{\ast}}^{v}(x_{1})\,
  {\phi}_{V}^{v}(x_{3})\,
   \Big[ -{\phi}_{D}^{p}(x_{2})\, m_{2}\,m_{c}\,u
   \nonumber \\ &+&
  {\phi}_{D}^{a}(x_{2})\, 4\,m_{1}^{2}\,p^{2}\,
  (x_{1}-\bar{x}_{2})
   \Big] \Big\}\, E_{n}(t^{C}_{c})\, {\alpha}_{s}(t^{C}_{c})\,
  C_{1}(t^{C}_{c})/N_{c}
   \label{amp:c-c-l},
   \end{eqnarray}
   \begin{eqnarray}
  {\cal M}^{C}_{c,N} &=&
   \frac{1}{N_{c}}\,
  {\int}_{0}^{1}dx_{1}
  {\int}_{0}^{1}dx_{2}
  {\int}_{0}^{1}dx_{3}
  {\int}_{0}^{\infty}db_{1}
  {\int}_{0}^{\infty}b_{2}db_{2}
  {\int}_{0}^{\infty}b_{3}db_{3}\,
  H^{C}_{n}({\alpha}^{C},{\beta}^{C}_{c},b_{2},b_{3})
   \nonumber \\ &{\times}&
  {\delta}(b_{1}-b_{3})\, E_{n}(t^{C}_{c})\,
  {\alpha}_{s}(t^{C}_{c}) \, \Big\{
  {\phi}_{B_{q}^{\ast}}^{V}(x_{1})\, {\phi}_{D}^{p}(x_{2})\,
  {\phi}_{V}^{V}(x_{3})\,2\,m_{1}\,m_{2}\,m_{3}\,m_{c}
   \nonumber \\ &+&
  {\phi}_{B_{q}^{\ast}}^{T}(x_{1})\, {\phi}_{D}^{a}(x_{2})\,
  {\phi}_{V}^{T}(x_{3})\,\Big[
  m_{1}^{2}\,s\,(\bar{x}_{2}-x_{1})
  +m_{3}^{2}\,t\,(x_{3}-\bar{x}_{2}) \Big] \Big\}\,
  C_{1}(t^{C}_{c})
   \label{amp:c-c-n},
   \end{eqnarray}
   \begin{eqnarray}
  {\cal M}^{C}_{c,T} &=&
   \frac{2}{N_{c}}\,
  {\int}_{0}^{1}dx_{1}
  {\int}_{0}^{1}dx_{2}
  {\int}_{0}^{1}dx_{3}
  {\int}_{0}^{\infty}db_{1}
  {\int}_{0}^{\infty}b_{2}db_{2}
  {\int}_{0}^{\infty}b_{3}db_{3}\,
  H^{C}_{n}({\alpha}^{C},{\beta}^{C}_{c},b_{2},b_{3})
   \nonumber \\ &{\times}&
  C_{1}(t^{C}_{c})\, \Big\{
  {\phi}_{B_{q}^{\ast}}^{T}(x_{1})\, {\phi}_{D}^{a}(x_{2})\,
  {\phi}_{V}^{T}(x_{3})\, \Big[ m_{1}^{2}\,(x_{1}-\bar{x}_{2})
  + m_{3}^{2}\,(\bar{x}_{2}-x_{3}) \Big]
   \nonumber \\ &-&
  {\phi}_{B_{q}^{\ast}}^{V}(x_{1})\, {\phi}_{D}^{p}(x_{2})\,
  {\phi}_{V}^{A}(x_{3})\,m_{2}\,m_{3}\,m_{c}/p
   \Big\}\, E_{n}(t^{C}_{c})\, {\alpha}_{s}(t^{C}_{c})\,
  {\delta}(b_{1}-b_{3})
   \label{amp:c-c-t},
   \end{eqnarray}
   \begin{eqnarray}
  {\cal M}^{C}_{d,P} &=&
   \frac{1}{N_{c}}\,
  {\int}_{0}^{1}dx_{1}
  {\int}_{0}^{1}dx_{2}
  {\int}_{0}^{1}dx_{3}
  {\int}_{0}^{\infty}db_{1}
  {\int}_{0}^{\infty}b_{2}db_{2}
  {\int}_{0}^{\infty}b_{3}db_{3}\,
  H^{C}_{n}({\alpha}^{C},{\beta}^{C}_{d},b_{2},b_{3})\,
  E_{n}(t^{C}_{d})
   \nonumber \\ &{\times}&
  {\delta}(b_{1}-b_{3})\, {\alpha}_{s}(t^{C}_{d})\,
  C_{1}(t^{C}_{d})\, {\phi}_{D}^{a}(x_{2})\, \Big\{
  {\phi}_{B_{q}^{\ast}}^{v}(x_{1})\, {\phi}_{P}^{a}(x_{3})\,
  2\,m_{1}\,p\,s\,(x_{2}-x_{3})
   \nonumber \\ &+&
  {\phi}_{B_{q}^{\ast}}^{t}(x_{1})\,m_{1}\,{\mu}_{P}\,
   \Big[ {\phi}_{P}^{p}(x_{3})\, 2\,m_{1}\,p\, (x_{3}-x_{1})
   + {\phi}_{P}^{t}(x_{3})\,(2\,m_{2}^{2}\,x_{2}+s\,x_{3}-t\,x_{1})
   \Big] \Big\}
   \label{amp:c-d-p},
   \end{eqnarray}
   \begin{eqnarray}
  {\cal M}^{C}_{d,L} &=&
   \frac{1}{N_{c}}\,
  {\int}_{0}^{1}dx_{1}
  {\int}_{0}^{1}dx_{2}
  {\int}_{0}^{1}dx_{3}
  {\int}_{0}^{\infty}db_{1}
  {\int}_{0}^{\infty}b_{2}db_{2}
  {\int}_{0}^{\infty}b_{3}db_{3}\,
  H^{C}_{n}({\alpha}^{C},{\beta}^{C}_{d},b_{2},b_{3})\,
  E_{n}(t^{C}_{d})
   \nonumber \\ &{\times}&
  {\delta}(b_{1}-b_{3})\, {\alpha}_{s}(t^{C}_{d})\,
  C_{1}(t^{C}_{d})\, {\phi}_{D}^{a}(x_{2})\, \Big\{
  {\phi}_{B_{q}^{\ast}}^{v}(x_{1})\, {\phi}_{V}^{v}(x_{3})\,
  4\,m_{1}^{2}\,p^{2}\,(x_{2}-x_{3})
   \nonumber \\ &+&
  {\phi}_{B_{q}^{\ast}}^{t}(x_{1})\,m_{1}\,m_{3}\,
   \Big[ {\phi}_{V}^{s}(x_{3})\, 2\,m_{1}\,p\, (x_{3}-x_{1})
   + {\phi}_{V}^{t}(x_{3})\,(2\,m_{2}^{2}\,x_{2}+s\,x_{3}-t\,x_{1})
   \Big] \Big\}
   \label{amp:c-d-l},
   \end{eqnarray}
   \begin{eqnarray}
  {\cal M}^{C}_{d,N} &=&
   \frac{1}{N_{c}}\,
  {\int}_{0}^{1}dx_{1}
  {\int}_{0}^{1}dx_{2}
  {\int}_{0}^{1}dx_{3}
  {\int}_{0}^{\infty}db_{1}
  {\int}_{0}^{\infty}b_{2}db_{2}
  {\int}_{0}^{\infty}b_{3}db_{3}\,
  H^{C}_{n}({\alpha}^{C},{\beta}^{C}_{d},b_{2},b_{3})\,
  E_{n}(t^{C}_{d})
   \nonumber \\ & & \!\!\!\! \!\!\!\! \!\!\!\!
  \!\!\!\! \!\!\!\! \!\!\!\!  {\times}\,
  {\delta}(b_{1}-b_{3})\,{\alpha}_{s}(t^{C}_{d})\,
  C_{1}(t^{C}_{d})\, {\phi}_{B_{q}^{\ast}}^{T}(x_{1})\,
  {\phi}_{D}^{a}(x_{2})\, {\phi}_{V}^{T}(x_{3})\,
   \Big\{ m_{1}^{2}\,s\,(x_{1}-x_{2})
   +m_{3}^{2}\,t\,(x_{2}-x_{3}) \Big\}
   \label{amp:c-d-n},
   \end{eqnarray}
   \begin{eqnarray}
  {\cal M}^{C}_{d,T} &=&
   \frac{2}{N_{c}}\,
  {\int}_{0}^{1}dx_{1}
  {\int}_{0}^{1}dx_{2}
  {\int}_{0}^{1}dx_{3}
  {\int}_{0}^{\infty}db_{1}
  {\int}_{0}^{\infty}b_{2}db_{2}
  {\int}_{0}^{\infty}b_{3}db_{3}\,
  H^{C}_{n}({\alpha}^{C},{\beta}^{C}_{d},b_{2},b_{3})\,
  E_{n}(t^{C}_{d})
   \nonumber \\ & & \!\!\!\! \!\!\!\! \!\!\!\!
  \!\!\!\! {\times}\,
  {\delta}(b_{1}-b_{3})\,{\alpha}_{s}(t^{C}_{d})\,
  C_{1}(t^{C}_{d})\, {\phi}_{B_{q}^{\ast}}^{T}(x_{1})\,
  {\phi}_{D}^{a}(x_{2})\, {\phi}_{V}^{T}(x_{3})\,
   \Big\{ m_{1}^{2}\,(x_{2}-x_{1})
   +m_{3}^{2}\,(x_{3}-x_{2}) \Big\}
   \label{amp:c-d-t}.
   \end{eqnarray}

  The functions $H_{f,n}^{C}$ have the similar expressions for
  $H_{f,n}^{T}$, i.e.,
   \begin{equation}
   H_{f}^{C}({\alpha},{\beta},b_{i},b_{j})\, =\,
   H_{f}^{T}({\alpha},{\beta},b_{i},b_{j})
   \label{amp:hfc},
   \end{equation}
   \begin{equation}
   H_{n}^{C}({\alpha},{\beta},b_{i},b_{j})\, =\,
   H_{n}^{T}({\alpha},{\beta},b_{i},b_{j})
   \label{amp:hnc}.
   \end{equation}

  The Sudakov factor $E_{f}^{C}$ are defined as
   \begin{equation}
   E_{f}^{C}(t)\ =\ {\exp}\{ -S_{B_{q}^{\ast}}(t)-S_{M}(t) \}
   \label{sudakov-fc},
   \end{equation}
  and the expressions for $E_{n}(t)$, $S_{B_{q}^{\ast}}(t)$,
  $S_{D}(t)$ and $S_{M}(t)$ are the same as those given in
  the Appendix \ref{block-t}.
  ${\alpha}^{C}$ and ${\beta}_{i}^{C}$ are the gluon and quark
  virtualities; the subscripts of ${\beta}_{i}^{C}$ and $t_{i}^{C}$
  correspond to the diagram indices of Fig.\ref{fig:fey-c}.
   \begin{eqnarray}
  {\alpha}^{C} &=& x_{1}^{2}\,m_{1}^{2}+x_{3}^{2}\,m_{3}^{2}-x_{1}\,x_{3}\,u
   \label{gluon-c}, \\
  {\beta}_{a}^{C} &=& x_{3}^{2}\,m_{3}^{2}-x_{3}\,u+m_{1}^{2}-m_{b}^{2}
   \label{beta-ca}, \\
  {\beta}_{b}^{C} &=& x_{1}^{2}\,m_{1}^{2}-x_{1}\,u+m_{3}^{2}
   \label{beta-cb}, \\
  {\beta}_{c}^{C} &=& {\alpha}^{C}+\bar{x}_{2}^{2}\,m_{2}^{2}
       -x_{1}\,\bar{x}_{2}\,t+x_{3}\,\bar{x}_{2}\,s-m_{c}^{2}
   \label{beta-cc}, \\
  {\beta}_{d}^{C} &=& {\alpha}^{C}+x_{2}^{2}\,m_{2}^{2}
       -x_{1}\,x_{2}\,t+x_{2}\,x_{3}\,s
   \label{beta-cd}, \\
   t_{a(b)}^{C} &=&
  {\max}(\sqrt{-{\alpha}^{C}},\sqrt{{\vert}{\beta}_{a(b)}^{C}{\vert}},1/b_{1},1/b_{3})
   \label{t-cab}, \\
   t_{c(d)}^{C} &=&
  {\max}(\sqrt{-{\alpha}^{C}},\sqrt{{\vert}{\beta}_{c(d)}^{C}{\vert}},1/b_{2},1/b_{3})
   \label{t-ccd}.
   \end{eqnarray}

  \section{Amplitude building blocks for the annihilation
   $\overline{B}^{{\ast}0}$ ${\to}$ $DM$ decays}
  \label{block-a}
  The expressions of the amplitude building blocks ${\cal M}^{A}_{i,j}$
  for the annihilation topologies are listed as follows, where the
  subscript $i$ corresponds to the diagram indices of Fig.\ref{fig:fey-a};
  and $j$ corresponds to different helicity amplitudes.
   \begin{eqnarray}
  {\cal M}^{A}_{a,P} &=&
  {\int}_{0}^{1}dx_{2}
  {\int}_{0}^{1}dx_{3}
  {\int}_{0}^{\infty}b_{2}db_{2}
  {\int}_{0}^{\infty}b_{3}db_{3}\,
  H^{A}_{f}({\alpha}^{A},{\beta}^{A}_{a},b_{2},b_{3})\,
  E^{A}_{f}(t^{A}_{a})\, {\alpha}_{s}(t^{A}_{a})
   \nonumber \\ &{\times}&
   a_{2}(t^{A}_{a})\,
   \Big\{ {\phi}_{D}^{p}(x_{2})\, \Big[
  {\phi}_{P}^{a}(x_{3})\,4\,m_{1}\,m_{2}\,m_{c}\,p
 +{\phi}_{P}^{p}(x_{3})\,4\,m_{1}\,m_{2}\,{\mu}_{P}\,p\,x_{3}
   \nonumber \\ &+&
  {\phi}_{P}^{t}(x_{3})\,2\,m_{2}\, {\mu}_{P}\,(t+u\,\bar{x}_{3})
   \Big] - {\phi}_{D}^{a}(x_{2})\, \Big[
  {\phi}_{P}^{p}(x_{3})\,\,2\,m_{1}\,m_{c}\,{\mu}_{P}\,p
   \nonumber \\ &+&
  {\phi}_{P}^{a}(x_{3})\,2\,m_{1}\,p\,
  (m_{1}^{2}\,\bar{x}_{3}+m_{2}^{2}\,x_{3})
 +{\phi}_{P}^{t}(x_{3})\,m_{c}\,{\mu}_{P}\,t \Big] \Big\}
   \label{amp:a-a-p},
   \end{eqnarray}
   \begin{eqnarray}
  {\cal M}^{A}_{a,L} &=&
  {\int}_{0}^{1}dx_{2}
  {\int}_{0}^{1}dx_{3}
  {\int}_{0}^{\infty}b_{2}db_{2}
  {\int}_{0}^{\infty}b_{3}db_{3}\,
  H^{A}_{f}({\alpha}^{A},{\beta}^{A}_{a},b_{2},b_{3})\,
  E^{A}_{f}(t^{A}_{a})\, {\alpha}_{s}(t^{A}_{a})
   \nonumber \\ &{\times}&
   a_{2}(t^{A}_{a})\,
   \Big\{ {\phi}_{D}^{p}(x_{2})\, \Big[
  {\phi}_{V}^{v}(x_{3})\,2\,m_{2}\,m_{c}\,u
 -{\phi}_{V}^{t}(x_{3})\,2\,m_{2}\,m_{3}\,(t+u\,\bar{x}_{3})
   \nonumber \\ &-&
  {\phi}_{V}^{s}(x_{3})\,4\,m_{1}\, m_{2}\,m_{3}\,p\,x_{3}
   \Big] + {\phi}_{D}^{a}(x_{2})\, \Big[
  {\phi}_{V}^{s}(x_{3})\,2\,m_{1}\,m_{3}\,m_{c}\,p
   \nonumber \\ &-&
  {\phi}_{V}^{v}(x_{3})\,
  (m_{2}^{2}\,u\,x_{3}+m_{1}^{2}\,s\,\bar{x}_{3})
 +{\phi}_{V}^{t}(x_{3})\,m_{3}\,m_{c}\,t \Big] \Big\}
   \label{amp:a-a-l},
   \end{eqnarray}
   \begin{eqnarray}
  {\cal M}^{A}_{a,N} &=&
  {\int}_{0}^{1}dx_{2}
  {\int}_{0}^{1}dx_{3}
  {\int}_{0}^{\infty}b_{2}db_{2}
  {\int}_{0}^{\infty}b_{3}db_{3}\,
  H^{A}_{f}({\alpha}^{A},{\beta}^{A}_{a},b_{2},b_{3})\,
  E^{A}_{f}(t^{A}_{a})
   \nonumber \\ &{\times}&
   \Big\{ {\phi}_{D}^{a}(x_{2})\, \Big[
  {\phi}_{V}^{V}(x_{3})\,m_{1}\,m_{3}\,(s\,\bar{x}_{3}+2\,m_{2}^{2})
 -{\phi}_{V}^{T}(x_{3})\,m_{1}\,m_{c}\,s
   \nonumber \\ &+&
  {\phi}_{V}^{A}(x_{3})\,2\,m_{1}^{2}\, m_{3}\,p\,\bar{x}_{3}
   \Big] - {\phi}_{D}^{p}(x_{2})\, \Big[
  {\phi}_{V}^{V}(x_{3})\,4\,m_{1}\,m_{2}\,m_{3}\,m_{c}
   \nonumber \\ &-&
   {\phi}_{V}^{T}(x_{3})\,2\,m_{1}\,m_{2}\,(s+2\,m_{3}^{2}\,\bar{x}_{3})
   \Big] \Big\}\, {\alpha}_{s}(t^{A}_{a})\, a_{2}(t^{A}_{a})
   \label{amp:a-a-n},
   \end{eqnarray}
   \begin{eqnarray}
  {\cal M}^{A}_{a,T} &=&
  {\int}_{0}^{1}dx_{2}
  {\int}_{0}^{1}dx_{3}
  {\int}_{0}^{\infty}b_{2}db_{2}
  {\int}_{0}^{\infty}b_{3}db_{3}\,
  H^{A}_{f}({\alpha}^{A},{\beta}^{A}_{a},b_{2},b_{3})\,
  E^{A}_{f}(t^{A}_{a})
   \nonumber \\ &{\times}&
   \Big\{ {\phi}_{D}^{p}(x_{2})\, 4\,m_{2}\Big[
  {\phi}_{V}^{T}(x_{3})\,m_{1}
 +{\phi}_{V}^{A}(x_{3})\,m_{3}\,m_{c}/p \Big]
   \nonumber \\ &-&
  {\phi}_{D}^{a}(x_{2})\, \Big[
  {\phi}_{V}^{V}(x_{3})\,2\,m_{1}\,m_{3}\,\bar{x}_{3}
 +{\phi}_{V}^{T}(x_{3})\,2\,m_{1}\,m_{c}
   \nonumber \\ &+&
  {\phi}_{V}^{A}(x_{3})\,(m_{3}/p)\,(s\,\bar{x}_{3}+2\,m_{2}^{2})
   \Big] \Big\}\, {\alpha}_{s}(t^{A}_{a}) \, a_{2}(t^{A}_{a})
   \label{amp:a-a-t},
   \end{eqnarray}
   \begin{eqnarray}
  {\cal M}^{A}_{b,P} &=&
  2\,m_{1}\,p\, {\int}_{0}^{1}dx_{2}
  {\int}_{0}^{1}dx_{3}
  {\int}_{0}^{\infty}b_{2}db_{2}
  {\int}_{0}^{\infty}b_{3}db_{3}\,
  H^{A}_{f}({\alpha}^{A},{\beta}^{A}_{b},b_{3},b_{2})\,
  E^{A}_{f}(t^{A}_{b})\, {\alpha}_{s}(t^{A}_{b})
   \nonumber \\ &{\times}&
   a_{2}(t^{A}_{b})\,
   \Big\{ {\phi}_{D}^{p}(x_{2})\, {\phi}_{P}^{p}(x_{3})\,
   2\,m_{2}\,{\mu}_{P}\,\bar{x}_{2}
   -{\phi}_{D}^{a}(x_{2})\, {\phi}_{P}^{a}(x_{3})\,
    (m_{1}^{2}\,x_{2}+m_{3}^{2}\,\bar{x}_{2}) \Big\}
   \label{amp:a-b-p},
   \end{eqnarray}
   \begin{eqnarray}
  {\cal M}^{A}_{b,L} &=&
  -{\int}_{0}^{1}dx_{2}
  {\int}_{0}^{1}dx_{3}
  {\int}_{0}^{\infty}b_{2}db_{2}
  {\int}_{0}^{\infty}b_{3}db_{3}\,
  H^{A}_{f}({\alpha}^{A},{\beta}^{A}_{b},b_{3},b_{2})\,
  E^{A}_{f}(t^{A}_{b})\, {\alpha}_{s}(t^{A}_{b})\, a_{2}(t^{A}_{b})
   \nonumber \\ &{\times}&
   \Big\{ {\phi}_{D}^{p}(x_{2})\, {\phi}_{V}^{s}(x_{3})\,
   4\,m_{1}\,m_{2}\,m_{3}\,p\,\bar{x}_{2}
   +{\phi}_{D}^{a}(x_{2})\, {\phi}_{V}^{v}(x_{3})\,
    (m_{1}^{2}\,s\,x_{2}+m_{3}^{2}\,t\,\bar{x}_{2}) \Big\}
   \label{amp:a-b-l},
   \end{eqnarray}
   \begin{eqnarray}
  {\cal M}^{A}_{b,N} &=&
  m_{1}\,m_{3}\,{\int}_{0}^{1}dx_{2}
  {\int}_{0}^{1}dx_{3}
  {\int}_{0}^{\infty}b_{2}db_{2}
  {\int}_{0}^{\infty}b_{3}db_{3}\,
  H^{A}_{f}({\alpha}^{A},{\beta}^{A}_{b},b_{3},b_{2})\,
  E^{A}_{f}(t^{A}_{b})
   \nonumber \\ &{\times}&
  {\alpha}_{s}(t^{A}_{b})\, a_{2}(t^{A}_{b})\,
  {\phi}_{D}^{a}(x_{2})\, \Big\{ {\phi}_{V}^{V}(x_{3})\,
  (s+2\,m_{2}^{2}\,x_{2})- {\phi}_{V}^{A}(x_{3})\,2\,m_{1}\,p \Big\}
   \label{amp:a-b-n},
   \end{eqnarray}
   \begin{eqnarray}
  {\cal M}^{A}_{b,T} &=&
  {\int}_{0}^{1}dx_{2}
  {\int}_{0}^{1}dx_{3}
  {\int}_{0}^{\infty}b_{2}db_{2}
  {\int}_{0}^{\infty}b_{3}db_{3}\,
  H^{A}_{f}({\alpha}^{A},{\beta}^{A}_{b},b_{3},b_{2})\,
  E^{A}_{f}(t^{A}_{b})\, {\alpha}_{s}(t^{A}_{b})
   \nonumber \\ &{\times}&
  a_{2}(t^{A}_{b})\,
  {\phi}_{D}^{a}(x_{2})\, \Big\{ {\phi}_{V}^{V}(x_{3})\,
  2\,m_{1}\,m_{3}- {\phi}_{V}^{A}(x_{3})\,(m_{3}/p)\,
   (s+2\,m_{2}^{2}\,x_{2}) \Big\}
   \label{amp:a-b-t},
   \end{eqnarray}
   \begin{eqnarray}
  {\cal M}^{A}_{c,P} &=&
  {\int}_{0}^{1}dx_{1}
  {\int}_{0}^{1}dx_{2}
  {\int}_{0}^{1}dx_{3}
  {\int}_{0}^{\infty}b_{1}db_{1}
  {\int}_{0}^{\infty}b_{2}db_{2}
  {\int}_{0}^{\infty}db_{3}\,
  H^{A}_{n}({\alpha}^{A},{\beta}^{A}_{c},b_{1},b_{2})
   \nonumber \\ &{\times}&
  {\delta}(b_{2}-b_{3}) \,
   \Big\{ {\phi}_{D}^{a}(x_{2})\,
  {\phi}_{P}^{a}(x_{3})\, 2\,m_{1}\,p \Big[
  {\phi}_{B_{q}^{\ast}}^{v}(x_{1})\,
  (s\,x_{2}+2\,m_{3}^{2}\,\bar{x}_{3}-u\,\bar{x}_{1})
   \nonumber \\ &+&
  {\phi}_{B_{q}^{\ast}}^{t}(x_{1})\,m_{1}\,m_{b} \Big] +
  {\phi}_{B_{q}^{\ast}}^{v}(x_{1})\, {\phi}_{D}^{p}(x_{2})\,
  m_{2}\,{\mu}_{p}\,\Big[ {\phi}_{P}^{p}(x_{3})\,
  2\,m_{1}\,p\,(x_{2}-\bar{x}_{3})
  \nonumber \\ &+&
  {\phi}_{P}^{t}(x_{3})\,(2\,m_{1}^{2}\,\bar{x}_{1}
  -t\,x_{2}-u\,\bar{x}_{3}) \Big] \Big\}\, E_{n}(t^{A}_{c})\,
  {\alpha}_{s}(t^{A}_{c})\, C_{1}(t^{A}_{c})/N_{c}
   \label{amp:a-c-p},
   \end{eqnarray}
   \begin{eqnarray}
  {\cal M}^{A}_{c,L} &=&
  {\int}_{0}^{1}dx_{1}
  {\int}_{0}^{1}dx_{2}
  {\int}_{0}^{1}dx_{3}
  {\int}_{0}^{\infty}b_{1}db_{1}
  {\int}_{0}^{\infty}b_{2}db_{2}
  {\int}_{0}^{\infty}db_{3}\,
  H^{A}_{n}({\alpha}^{A},{\beta}^{A}_{c},b_{1},b_{2})
   \nonumber \\ &{\times}&
  {\delta}(b_{2}-b_{3})\,
   \Big\{ {\phi}_{B_{q}^{\ast}}^{v}(x_{1})\,
  {\phi}_{D}^{p}(x_{2})\, m_{2}\,m_{3} \Big[
  {\phi}_{V}^{t}(x_{3})\,
  (t\,x_{2}+u\,\bar{x}_{3}-2\,m_{1}^{2}\,\bar{x}_{1})
   \nonumber \\ &+&
  {\phi}_{V}^{s}(x_{3})\, 2\,m_{1}\,p\,(\bar{x}_{3}-x_{2})
   \Big] + {\phi}_{D}^{a}(x_{2})\, {\phi}_{V}^{v}(x_{3})\,
   \Big[ {\phi}_{B_{q}^{\ast}}^{t}(x_{1})\, m_{1}\,m_{b}\,s
   \nonumber \\ &+&
  {\phi}_{B_{q}^{\ast}}^{v}(x_{1})\,
   4\,m_{1}^{2}\,p^{2}\,(x_{2}-\bar{x}_{1}) \Big]
   \Big\}\, E_{n}(t^{A}_{c})\,
  {\alpha}_{s}(t^{A}_{c})\, C_{1}(t^{A}_{c})/N_{c}
   \label{amp:a-c-l},
   \end{eqnarray}
   \begin{eqnarray}
  {\cal M}^{A}_{c,N} &=&
  {\int}_{0}^{1}dx_{1}
  {\int}_{0}^{1}dx_{2}
  {\int}_{0}^{1}dx_{3}
  {\int}_{0}^{\infty}b_{1}db_{1}
  {\int}_{0}^{\infty}b_{2}db_{2}
  {\int}_{0}^{\infty}db_{3}\,
  H^{A}_{n}({\alpha}^{A},{\beta}^{A}_{c},b_{1},b_{2})\,
  E_{n}(t^{A}_{c})
   \nonumber \\ &{\times}&
  {\delta}(b_{2}-b_{3})\, {\alpha}_{s}(t^{A}_{c})\,
   \Big\{ {\phi}_{B_{q}^{\ast}}^{V}(x_{1})\,
  {\phi}_{D}^{p}(x_{2})\, {\phi}_{V}^{T}(x_{3})\,
  m_{1}\,m_{2}\,(u\,\bar{x}_{1}-s\,x_{2}-2\,m_{3}^{2}\,\bar{x}_{3})
   \nonumber \\ &+&
  {\phi}_{B_{q}^{\ast}}^{T}(x_{1})\, {\phi}_{D}^{a}(x_{2})\,
  m_{3}\,m_{b}\, \Big[ {\phi}_{V}^{A}(x_{3})\,2\,m_{1}\,p
  -{\phi}_{V}^{V}(x_{3})\,t  \Big]
   \Big\}\, C_{1}(t^{A}_{c})/N_{c}
   \label{amp:a-c-n},
   \end{eqnarray}
   \begin{eqnarray}
  {\cal M}^{A}_{c,T} &=&
  {\int}_{0}^{1}dx_{1}
  {\int}_{0}^{1}dx_{2}
  {\int}_{0}^{1}dx_{3}
  {\int}_{0}^{\infty}b_{1}db_{1}
  {\int}_{0}^{\infty}b_{2}db_{2}
  {\int}_{0}^{\infty}db_{3}\,
  H^{A}_{n}({\alpha}^{A},{\beta}^{A}_{c},b_{1},b_{2})
   \nonumber \\ &{\times}&
  {\delta}(b_{2}-b_{3}) \,
  E_{n}(t^{A}_{c})\, {\alpha}_{s}(t^{A}_{c})\,
   \Big\{ {\phi}_{B_{q}^{\ast}}^{V}(x_{1})\,
  {\phi}_{D}^{p}(x_{2})\, {\phi}_{V}^{T}(x_{3})\,
  2\,m_{1}\,m_{2}\,(\bar{x}_{1}-x_{2})
   \nonumber \\ &+&
  {\phi}_{B_{q}^{\ast}}^{T}(x_{1})\, {\phi}_{D}^{a}(x_{2})\,
  m_{3}\,m_{b}\, \Big[ {\phi}_{V}^{A}(x_{3})\,t/(m_{1}\,p)
  -2\,{\phi}_{V}^{V}(x_{3})  \Big]
   \Big\}\, C_{1}(t^{A}_{c})/N_{c}
   \label{amp:a-c-t},
   \end{eqnarray}
   \begin{eqnarray}
  {\cal M}^{A}_{d,P} &=&
   \frac{1}{N_{c}}\,
  {\int}_{0}^{1}dx_{1}
  {\int}_{0}^{1}dx_{2}
  {\int}_{0}^{1}dx_{3}
  {\int}_{0}^{\infty}b_{1}db_{1}
  {\int}_{0}^{\infty}b_{2}db_{2}
  {\int}_{0}^{\infty}db_{3}\,
  H^{A}_{n}({\alpha}^{A},{\beta}^{A}_{d},b_{1},b_{2})\, E_{n}(t^{A}_{d})
   \nonumber \\ &{\times}&
  {\delta}(b_{2}-b_{3})\, {\alpha}_{s}(t^{A}_{d})\,
  C_{1}(t^{A}_{d})\, {\phi}_{B_{q}^{\ast}}^{v}(x_{1})\, \Big\{
  {\phi}_{D}^{a}(x_{2})\, {\phi}_{P}^{a}(x_{3})\,
  2\,m_{1}\,p\,( 2\,m_{2}^{2}\,x_{2}+s\,\bar{x}_{3}-t\,x_{1})
   \nonumber \\ &+&
  {\phi}_{D}^{p}(x_{2})\,
  m_{2}\,{\mu}_{P}\, \Big[ {\phi}_{P}^{t}(x_{3})\,
  (2\,m_{1}^{2}\,x_{1}-t\,x_{2}-u\,\bar{x}_{3})
  +{\phi}_{P}^{p}(x_{3})\,2\,m_{1}\,p\,(\bar{x}_{3}-x_{2})
   \Big] \Big\}
   \label{amp:a-d-p},
   \end{eqnarray}
   \begin{eqnarray}
  {\cal M}^{A}_{d,L} &=&
   \frac{1}{N_{c}}\,
  {\int}_{0}^{1}dx_{1}
  {\int}_{0}^{1}dx_{2}
  {\int}_{0}^{1}dx_{3}
  {\int}_{0}^{\infty}b_{1}db_{1}
  {\int}_{0}^{\infty}b_{2}db_{2}
  {\int}_{0}^{\infty}db_{3}\,
  H^{A}_{n}({\alpha}^{A},{\beta}^{A}_{d},b_{1},b_{2})\, E_{n}(t^{A}_{d})
   \nonumber \\ &{\times}&
  {\delta}(b_{2}-b_{3})\, {\alpha}_{s}(t^{A}_{d})\,
  C_{1}(t^{A}_{d})\, {\phi}_{B_{q}^{\ast}}^{v}(x_{1})\, \Big\{
  {\phi}_{D}^{a}(x_{2})\, {\phi}_{V}^{v}(x_{3})\,
  u\,( 2\,m_{2}^{2}\,x_{2}+s\,\bar{x}_{3}-t\,x_{1})
   \nonumber \\ &-&
  {\phi}_{D}^{p}(x_{2})\,
  m_{2}\,m_{3}\, \Big[ {\phi}_{V}^{t}(x_{3})\,
  (2\,m_{1}^{2}\,x_{1}-t\,x_{2}-u\,\bar{x}_{3})
  +{\phi}_{V}^{s}(x_{3})\,2\,m_{1}\,p\,(\bar{x}_{3}-x_{2})
   \Big] \Big\}
   \label{amp:a-d-l},
   \end{eqnarray}
   \begin{eqnarray}
  {\cal M}^{A}_{d,N} &=&
   \frac{1}{N_{c}}\,
  {\int}_{0}^{1}dx_{1}
  {\int}_{0}^{1}dx_{2}
  {\int}_{0}^{1}dx_{3}
  {\int}_{0}^{\infty}b_{1}db_{1}
  {\int}_{0}^{\infty}b_{2}db_{2}
  {\int}_{0}^{\infty}db_{3}\,
  H^{A}_{n}({\alpha}^{A},{\beta}^{A}_{d},b_{1},b_{2})
   \nonumber \\ &{\times}&
  E_{n}(t^{A}_{d})\, {\alpha}_{s}(t^{A}_{d})\,
  {\phi}_{B_{q}^{\ast}}^{V}(x_{1})\, \Big\{
  {\phi}_{D}^{a}(x_{2})\, {\phi}_{V}^{V}(x_{3})\,
  2\,m_{1}\,m_{3}\,( t\,x_{1}-2\,m_{2}^{2}\,x_{2}-s\,\bar{x}_{3})
   \nonumber \\ &+&
  {\phi}_{D}^{p}(x_{2})\,{\phi}_{V}^{T}(x_{3})\,
  m_{1}\,m_{2}\,(u\,x_{1}-s\,x_{2}-2\,m_{3}^{2}\,\bar{x}_{3})
   \Big] \Big\}\, C_{1}(t^{A}_{d})\, {\delta}(b_{2}-b_{3})
   \label{amp:a-d-n},
   \end{eqnarray}
   \begin{eqnarray}
  {\cal M}^{A}_{d,T} &=&
   \frac{2}{N_{c}}\,
  {\int}_{0}^{1}dx_{1}
  {\int}_{0}^{1}dx_{2}
  {\int}_{0}^{1}dx_{3}
  {\int}_{0}^{\infty}b_{1}db_{1}
  {\int}_{0}^{\infty}b_{2}db_{2}
  {\int}_{0}^{\infty}db_{3}\,
  H^{A}_{n}({\alpha}^{A},{\beta}^{A}_{d},b_{1},b_{2})
   \nonumber \\ &{\times}&
  E_{n}(t^{A}_{d})\, {\alpha}_{s}(t^{A}_{d})\,
   {\phi}_{B_{q}^{\ast}}^{V}(x_{1})\, \Big\{
  {\phi}_{D}^{a}(x_{2})\, {\phi}_{V}^{A}(x_{3})\,
  (m_{3}/p)\,( 2\,m_{2}^{2}\,x_{2}+s\,\bar{x}_{3}-t\,x_{1})
   \nonumber \\ &+&
  {\phi}_{D}^{p}(x_{2})\,{\phi}_{V}^{T}(x_{3})\,
  m_{1}\,m_{2}\,(x_{1}-x_{2}) \Big] \Big\}\,
  C_{1}(t^{A}_{d})\, {\delta}(b_{2}-b_{3})
   \label{amp:a-d-t}.
   \end{eqnarray}

  The functions $H_{f,n}^{A}$ and the Sudakov factor $E_{f}^{A}$
  are defined as follows.
   \begin{eqnarray}
   H_{f}^{A}({\alpha},{\beta},b_{i},b_{j}) &=&
   \frac{{\pi}^{2}}{4} \, \Big\{ i\,J_{0}(b_{i}\sqrt{{\alpha}})
   - Y_{0}(b_{i}\sqrt{{\alpha}}) \Big\}
   \nonumber \\ & & \!\!\!\! \!\!\!\! \!\!\!\! {\times}\,
   \Big\{ {\theta}(b_{i}-b_{j}) \Big[ i\,J_{0}(b_{i}\sqrt{{\beta}})
   - Y_{0}(b_{i}\sqrt{{\beta}}) \Big] J_{0}(b_{j}\sqrt{{\beta}})
  +\, (b_{i} {\leftrightarrow} b_{j}) \Big\}
   \label{amp:hfa},
   \end{eqnarray}
   \begin{eqnarray}
   H_{n}^{A}({\alpha},{\beta},b_{i},b_{j}) &=&
    \Big\{ {\theta}(-{\beta})\, K_{0}(b_{i}\sqrt{-{\beta}})
  + \frac{\pi}{2}{\theta}(+{\beta})\,
    \Big[ i\,J_{0}(b_{i}\sqrt{{\beta}})
   - Y_{0}(b_{i}\sqrt{{\beta}}) \Big] \Big\}
   \nonumber \\ & & \!\!\!\! \!\!\!\! \!\!\!
  {\times}\, \frac{\pi}{2} \,
   \Big\{ {\theta}(b_{i}-b_{j}) \Big[ i\,J_{0}(b_{i}\sqrt{{\alpha}})
   - Y_{0}(b_{i}\sqrt{{\alpha}}) \Big]
   J_{0}(b_{j}\sqrt{{\alpha}})
   + (b_{i} {\leftrightarrow} b_{j}) \Big\}
   \label{amp:hna},
   \end{eqnarray}
   \begin{equation}
   E_{f}^{A}(t)\ =\ {\exp}\{ -S_{D}(t)-S_{M}(t) \}
   \label{sudakov-fa},
   \end{equation}
  and the expressions for $E_{n}(t)$, $S_{B_{q}^{\ast}}(t)$,
  $S_{D}(t)$ and $S_{M}(t)$ are the same as those given in
  the Appendix \ref{block-t}.
  ${\alpha}^{A}$ and ${\beta}_{i}^{A}$ are the gluon and quark
  virtualities; the subscripts of ${\beta}_{i}^{A}$ and $t_{i}^{A}$
  correspond to the diagram indices of Fig.\ref{fig:fey-a}.
   \begin{eqnarray}
  {\alpha}^{A} &=& x_{2}^{2}\,m_{2}^{2}+\bar{x}_{3}^{2}\,m_{3}^{2}+x_{2}\,\bar{x}_{3}\,s
   \label{gluon-a}, \\
  {\beta}_{a}^{A} &=& \bar{x}_{3}^{2}\,m_{3}^{2}+\bar{x}_{3}\,s+m_{2}^{2}-m_{c}^{2}
   \label{beta-aa}, \\
  {\beta}_{b}^{A} &=& x_{2}^{2}\,m_{2}^{2}+x_{2}\,s+m_{3}^{2}
   \label{beta-ab}, \\
  {\beta}_{c}^{A} &=& {\alpha}^{A}+\bar{x}_{1}^{2}\,m_{1}^{2}
                      -\bar{x}_{1}\,x_{2}\,t-\bar{x}_{1}\,\bar{x}_{3}\,u-m_{b}^{2}
   \label{beta-ac}, \\
  {\beta}_{d}^{A} &=& {\alpha}^{A}+x_{1}^{2}\,m_{1}^{2}
                      -x_{1}\,x_{2}\,t-x_{1}\,\bar{x}_{3}\,u
   \label{beta-ad}, \\
   t_{a(b)}^{A} &=&
  {\max}(\sqrt{{\alpha}^{A}},\sqrt{{\vert}{\beta}_{a(b)}^{A}{\vert}},1/b_{2},1/b_{3})
   \label{t-aab}, \\
   t_{c(d)}^{A} &=&
  {\max}(\sqrt{{\alpha}^{A}},\sqrt{{\vert}{\beta}_{c(d)}^{A}{\vert}},1/b_{1},1/b_{2})
   \label{t-acd}.
   \end{eqnarray}
   \end{appendix}

  
  \end{document}